\documentclass[structabstract]{aa}
\usepackage{graphicx}
\usepackage{txfonts}
\usepackage{natbib}

\usepackage[T1]{fontenc}

\bibpunct{(}{)}{;}{a}{}{,} % to follow the A&A style

\def\gtrsim{\mathrel{\hbox{\rlap{\hbox{\lower4pt\hbox{$\sim$}}}\hbox{$>$}}}}
\def\lesssim{\mathrel{\hbox{\rlap{\hbox{\lower4pt\hbox{$\sim$}}}\hbox{$<$}}}}
\newcommand{\hi}{H{\sc i} }

\newcommand{\msun}{M$_{\odot}$$\,$}

\newcommand{\arsec}{$^{\prime\prime}\!$}

\newcommand{\armin}{$^{\prime}\!\!$ }
\newcommand{\arminsp}{$^{\prime}\;$}
\newcommand{\vi}{$(V-I)_0$ }
\newcommand{\Z}{[M/H]}
\newcommand{\rsb}{R$_{25}$}

\begin{document}

   \title{Stellar structures in the outer regions of M33\thanks{Based on data collected at Subaru Telescope, which is operated by the National Astronomical Observatory of Japan.}}

   \subtitle{}

   \author{M.Grossi
          \inst{1}
          \and
          N. Hwang\inst{2}%\fnmsep\thanks{}
          \and E. Corbelli\inst{3}
          \and C. Giovanardi\inst{3}
          \and S. Okamoto\inst{4}
          \and N. Arimoto\inst{2,5}}
%          \and L.K. Hunt\inst{4}
%          \and L. Magrini\inst{4}
%          \and F. Palla\inst{4}}

\institute{
CAAUL, Observat\'orio Astron\'omico de Lisboa, Universidade de Lisboa, Tapada da Ajuda, 1349-018, Lisboa, Portugal\\
\email{grossi@oal.ul.pt}
\and
National Astronomical Observatory of Japan, Mitaka, Tokyo 181-8588, Japan
\and
INAF-Osservatorio Astrofisico di Arcetri, Largo Enrico Fermi 5, 50125 Firenze, Italy
\and
Kavli Institute for Astronomy and Astrophysics, Peking University, Yi He Yuan Lu 5, Hai Dian Qu, Beijing 100871, P. R. China
\and
Department of Astronomical Science, Graduate University of Advanced Studies, Mitaka, Tokyo 181-8588, Japan }

   \date{}

% \abstract{}{}{}{}{}
% 5 {} token are mandatory

  \abstract{}
{We present Subaru/Suprime-Cam deep $V$ and $I$ imaging of seven fields in the outer regions of M33. Our aim is to search for stellar structures corresponding to extended \hi clouds found in a recent 21-cm survey of the galaxy. Three fields probe a large \hi complex to the southeastern (SE) side of the galaxy. An additional three fields cover the northwestern (NW) side of the galaxy along the \hi warp. A final target field was chosen further north, at a projected distance of approximately 25 kpc, to study part of the large stellar plume recently discovered around M33.}
{We analyse the stellar population at $R >$ 10 kpc by means of $V$, $I$ colour magnitude diagrams reaching the red clump. We constrain the age and metallicity of the different stellar populations, search for density enhancements that correspond to the \hi features, and investigate the radial surface distribution of the stars. }
{We find evolved stellar populations in all fields out to 120\arminsp ($\sim$ 30 kpc),
while a diffuse population of young stars  ($\sim$ 200 Myr) is detected out to a galactocentric radius of 15 kpc.  The mean metallicity in the southern fields remains approximately constant at \Z = -0.7 beyond the edge of the optical disc,  from 40\arminsp out to 80\armin.
Along the northern fields probing the outer \hi disc, we also find a metallicity of \Z=-0.7 between 35\arminsp and 70\arminsp from the centre, which decreases to \Z = -1.0 at larger angular radii out to 120\armin.
In the northernmost field, outside the disc extent, the stellar population of the large stellar plume possibly related to a M33-M31 interaction is on average more metal-poor (\Z=-1.3) and older ($\gtrsim$ 6 Gyr).}
  % conclusions heading (mandatory)
{An exponential disc with a large scale-length ($\sim$ 7 kpc) fits well
the average distribution of stars detected in both the SE and NW regions from a galactocentric distance of 11 kpc out to 30~kpc.
The stellar disc extends beyond the \hi disc. The stellar distribution  at large radii is disturbed and,
although there is no clear correlation between the stellar
substructures and the location of the \hi clouds, this gives evidence of tidal interaction or accretion events.}%


   \maketitle

\section{Introduction}

Hierarchical models of structure formation predict that galaxies form after a series of accretion events involving lower mass systems \citep{2003ApJ...591..499A,2007MNRAS.374.1479G}. Within this scenario, a massive spiral galaxy such as the Milky Way should have accreted 100-200 satellites during its evolutionary history \citep{2005ApJ...635..931B}.
%The process of tidal disruption and accretion of these systems
This process
would leave its imprint in a diffuse stellar halo extending out to 10-100 times the optical radius of the galactic disc. The density, luminosity, and metallicity of the stellar halo can thus provide direct information about the history and evolution of a galaxy, and enable us to test hierarchical formation scenarios.

The discovery of the Sagittarius stream in the Milky Way (MW) revealed for the first time the role of satellite accretion in the build-up of a galaxy mass \citep{1994Natur.370..194I}.
Studies of the MW halo using data from the Sloan Digitised Sky Survey \citep[SDSS;][]{2000AJ....120.1579Y} revealed other tidal streams around our galaxy, proving that these features are relatively common and can survive %and maintain coherence in the galactic halo
for a few Gyrs \citep{2006ApJ...642L.137B,2006ApJ...645L..37G,2007ApJ...658..337B,2009ApJ...693.1118G}.

%(INT, Mega-CAM, PANDAs)
Similar features were discovered around M31, the other massive spiral of the Local Group (LG), that implied this galaxy has experienced a more violent merger history than the MW \citep{2006AJ....131.1436F,2008ApJ...682L..33F}. A giant stellar stream with a neutral hydrogen (H{\sc i}) counterpart was found in this galaxy \citep{1977MNRAS.181..573N,2001Natur.412...49I,2002AJ....124.1452F,2004ApJ...601L..39T}.
Much effort has been devoted to obtaining deeper optical images of the
halo of M31 with the Pan-Andromeda Archaeological Survey \citep[PAndAS;][]{2009Natur.461...66M}, which
revealed a large-scale metal-poor stellar distribution (stretching out to 150 kpc), and
uncovered additional extended substructures in the halo similar to the giant southern stream.
%%%%%%%%%%%%%%%%%%%%%%%%%%%%%%%%%%%%%%%%%%%%%%%%%%%%%%%%%%%%%%%%%%%%%%%%%%%%%%%%%%%%%%%%%%%%%%%%%%%%%%%%%%%%%%%%%%%%%%%
%%
%%
%%
%%%%%%%%%%%%%%%%%%%%%%%%%%%%%%%%%%%%%%%%%%%%%%%%%%%%%%%%%%%%%%%%%%%%%%%%%%%%%%%%%%%%%%%%%%%%%%%%%%%%%%%%%%%%%%%%%%%%%%%

While M31 clearly shows signs of disturbance in its spheroid and outer disc, M33, its brightest satellite and the third most massive member of the LG, has long been considered an example of a ``pure disc'' system %without a diffuse stellar spheroid
evolving in relative isolation, because of the lack of a prominent bulge, a seemingly unperturbed stellar disc, %\citep[][]{2007ApJ...669..315C},
and the absence of known nearby satellites.
The origin of bulgeless galaxies such as M33 is difficult to explain within the hierarchical structure formation scenario \citep{2001MNRAS.327.1334V}, because mergers of galaxies are expected to form a bulge component.
The inflow of gas at later times ($z > 1$) from the local environment is thought to play a major role in the growth of the discs of these systems \citep{2006MNRAS.368....2D,2009ApJ...694..396B}.
Nonetheless, the presence of a diffuse  faint stellar halo even in disc galaxies seems to be expected within the
framework of current galaxy formation scenarios, predicting that a diffuse stellar halo rather than a centrally-concentrated bulge would result after the late major-merger event
experienced by these galaxies    %. from the "scrambled trajectories of stars"
\citep{2010arXiv1010.1004B}.
%.

Studies of the
outskirts of M33 have started providing evidence of an extended stellar distribution beyond the optical disc, although it remains unclear whether this component corresponds to a stellar halo or an extended disc.
\citet{2004AJ....128..237B} derived a radial density stellar profile out to 1 degree from the centre of M33
%($r_{proj} \sim$ 16 kpc)
finding a clear dropoff in the star counts as a function of radius
%with a power law
($\propto R^{-1.46}$), that they interpreted as a evidence of a stellar spheroid component.  The metallicity peak of this sparse stellar population, [Fe/H] = -1.24 $\pm$ 0.04, was lower than the abundance in the disc ([Fe/H] $\sim$ -0.9), and comparable to the mean metallicity of a sample of halo globular clusters \citep{2000AJ....120.2437S}.
In a spectroscopic survey of red giant branch (RGB) stars in M33, \citet{2006ApJ...647L..25M} %provided also evidence for a stellar halo.
found that the stellar velocity distributions could be described by three different components:
the galaxy disc, a halo (with [Fe/H] = -1.5), and a third distribution possibly associated with a stellar stream.
Finally, the PAndAS survey revealed an extended stellar distribution around M33 well beyond the main optical disc out to 3 degrees (or a projected radius of $\sim$ 45 kpc) from its centre \citep{2009Natur.461...66M,2010ApJ...723.1038M}. This structure could either be the stellar debris of an accreted satellite, or the result of a tidal interaction with M31. This latter interpretation would be supported by the similar spatial orientation of the stellar features and the M33 \hi warp. The origin of the \hi warp remains disputed and previous studies have proposed that it is the result of an interaction with its massive neighbour \citep{1997ApJ...479..244C,2008MNRAS.390L..24B}.
Simulations of the relative motion of the M31/M33 system \citep{2005ApJ...633..894L,2009ApJ...703.1486P,2009Natur.461...66M},
converge on the possibility of a close encounter between the two galaxies occurring a few Gyr ago with a pericenter distance around or greater than 40 kpc, which could strip material from the M33 outer disc without severely distorting or disrupting the bulk of the stellar disc.

Additional information about the outer regions of M33 can be inferred from neutral hydrogen studies.
A 21-cm survey of the \hi distribution in a region of $3^{\circ} \times 3^{\circ}$ around M33 with
the Arecibo telescope \citep[ALFALFA;][]{2005AJ....130.2598G}
led to the finding of a population of \hi clouds in the environment of M33 \citep{2008A&A...487..161G}.
These clouds, with \hi masses ranging between $10^4$ and few times $10^6$ \msun \citep[at a distance of 840 kpc;][]{2001ApJ...553...47F}, were found within a projected distance of about 20 kpc, and they appear
to be distributed (in projection) along the major axis of the \hi disc, towards the direction of
M31.
These features could also be evidence of either a tidal interaction with M31, %which is now falling back onto the disc of M33,
or the accretion of a dwarf companion of M33.
In either case one might expect to find
 features in the stellar distribution that correspond to these \hi structures. The search for optical counterparts to the most massive and extended clouds may provide further insight into their origin, and the understanding of the outskirts of M33.

%%%%%%%%%%%%%%%%%%%%%%%%%%%%%%%%%%%%%%%%%%%%%%%%%%%%%%%%%%%%%%%%%%%%%%%%%%%%%%%%%%%%%%%%%%%%%%%%%%%%%%%%%%%%%%%%%%%%%%
%%%%%%%%%%%%%%%%%%%%%%%%%%%%%%%%%%%%%%%%%%%%%%%%%%%%%%%%%%%%%%%%%%%%%%%%%%%%%%%%%%%%%%%%%%%%%%%%%%%%%%%%%%%%%%%%%%%%%%
%%%%%%%%%%%%%%%%%%%%%%%%%%%%%%%%%%%%%%%%%%%%%%%%%%%%%%%%%%%%%%%%%%%%%%%%%%%%%%%%%%%%%%%%%%%%%%%%%%%%%%%%%%%%%%%%%%%%%%
%%%%%%%%%%%%%%%%%%%%%%%%%%%%%%%%%%%%%%%%%%%%%%%%%%%%%%%%%%%%%%%%%%%%%%%%%%%%%%%%%%%%%%%%%%%%%%%%%%%%%%%%%%%%%%%%%%%%%%

Taking advantage of the unique sensitivity and large field of view of the Subaru
Suprime-Cam instrument we mapped through the $V$ and $I$ filters sixregions in the southeastern (SE) and northwestern (NW) outer disc of M33 corresponding to the main \hi structures found in \citet{2008A&A...487..161G}. An additional field to the north of the galaxy %at a projected distance of about 30 kpc
was observed to study in more detail the nature of the stellar structure at very large radii found by \citet{2009Natur.461...66M}.
We used colour-magnitude diagrams (CMDs) reaching down to the red clump (RC) to constrain the age and the metallicity of the stars at large galactocentric distances, and we
searched for stellar density enhancements related to the gaseous features.

The paper is organised as follows: in Sect. 2, we describe the observations and data reduction process; in Sect. 3, we discuss the main properties of the CMDs of the target fields;
in Sect. 4, we derive the distance to M33 from the tip of the RGB.
We estimate the age and metallicity of the stellar populations in Sect. 5 from the comparison to evolutionary stellar population models, we derive in Sect. 6 the star formation history of the northernmost field of our survey using the CMD fitting method, while in Sect. 7 we show the spatial distribution of the stellar structures in the target regions. Discussion of our results is given in Sect. 8, and in Sect. 9 we summarize our conclusions.

   \begin{figure}
\includegraphics[width=8.8cm, bb= 20 0 480 450]{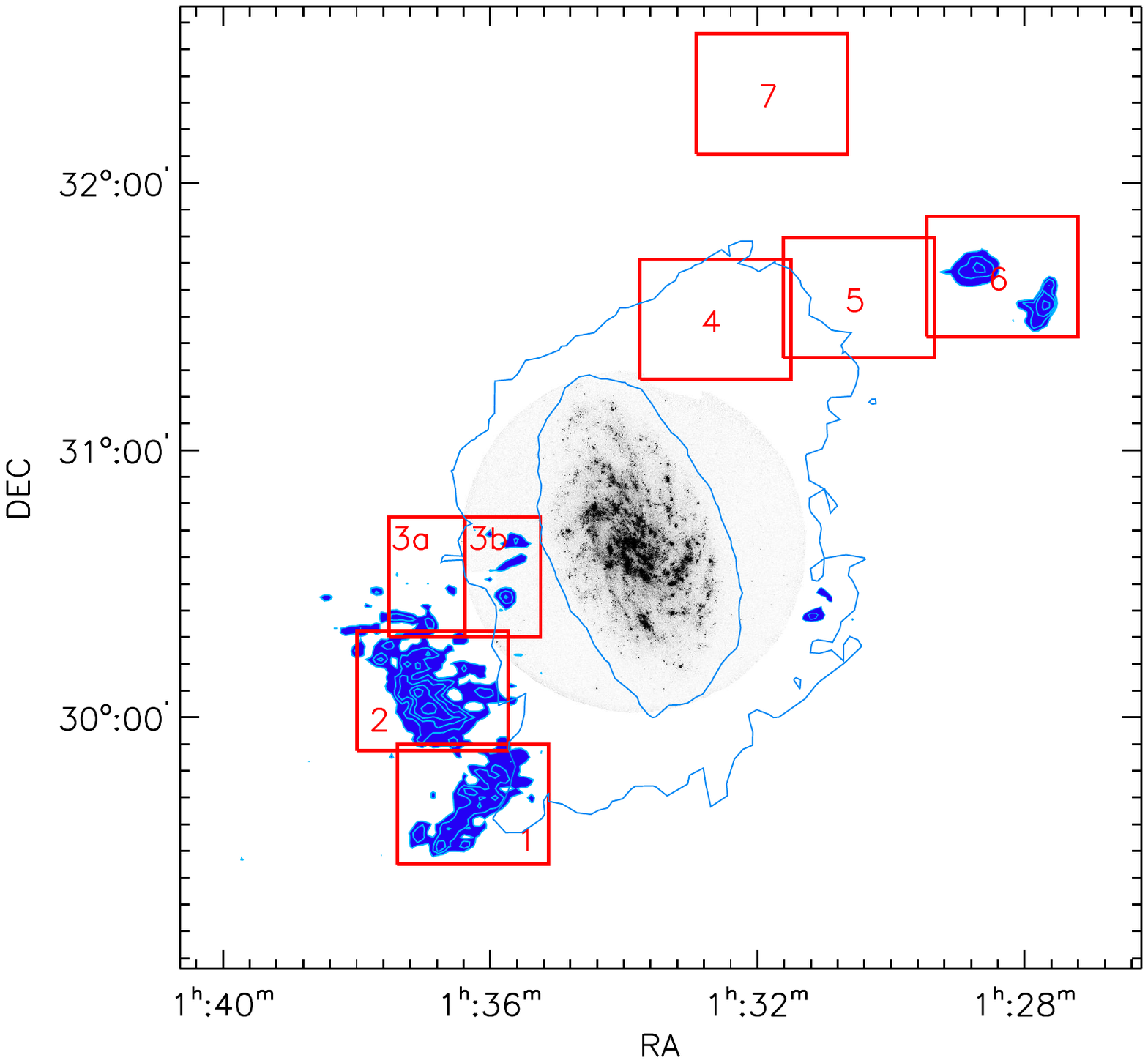}
   \caption{Location of the seven target fields around M33 imaged with Subaru/Suprime-Cam. The  M33 \hi disc is highlighted by two contours with a column density of N$_{HI}= 5$ and $50 \times 10^{19}$ cm$^{-2}$, respectively. Contours are overlaid on the GALEX (far ultraviolet) image of M33 \citep{2005ApJ...619L..67T}. The main \hi features to the SE and NW of the galaxy are also shown.}
\label{fields}
   \end{figure}

%__________________________________________________________________

\section{Observations and data reduction}

We obtained $V$ and $I$ images  with Suprime-Cam \citep{2002PASJ...54..833M} on the 8.2 meter Subaru telescope on the nights of August 21 and September 21-22, 2009. Suprime-Cam consists of ten CCDs of 2048$\times$4096 pixels with a pixel size of 0.2 arcsec/px and a total field of view of 34$\times$27 arcmin. The regions observed during the run are shown in Fig. \ref{fields}. The fields are labelled from one to seven and are related to various \hi features around the galaxy. Fields SE1-SE3 cover the large gaseous complex to the south-east of the galaxy (see Fig.1), whereas fields NW4-NW6 extend from the northern edge of the disc along the \hi warp to the northernmost \hi clouds detected around the galaxy in \citet{2008A&A...487..161G}. Field NW7 was chosen to analyse the stellar population of the large stellar feature %extended substructure at very large galactocentric distances that has been
uncovered by the PAndAS Survey. Finally, a field at approximately 4 degrees to the east of the optical center of M33 was selected to inspect the contamination from foreground and background sources.
For each field, we obtained a set of 5$\times$440s and 15$\times$320s exposures through the $V$ and $I$ filters, with total exposure times of 2200s and 4800s, respectively.
To fill the gaps between the CCD chips and improve the removal of the effects of cosmic rays and bad pixels, the individual exposures were dithered by 60 arcsec along a five position pattern.
Table 1 lists the coordinates of the eight target fields and the corresponding projected distance from the centre of M33.

%________________________________________________________________

\begin{table}
\caption{Central coordinates of the target fields observed with Subaru/Suprime-Cam, corresponding projected distance from M33 (for $d_{M33} =$ 840 kpc), seeing of the combined $V$ and $I$ images, and extinction values adopted for each field.}
\begin{center}
\begin{tabular}{lccrcc}
\hline \hline
Field & RA      & DEC     & $r$ & Seeing  & A$_V$ \\ %& A$_I$\\
      & (J2000) & (J2000) & kpc        & $V\:\:$/$\:\:I$ &  mag\\%   &  mag\\
\hline \hline
SE1     & 01:36:15.6 & 29:40:30 & 16.3 & 0\arsec.9/0\arsec.8 & 0.157 \\% & 0.092\\
SE2     & 01:36:51.6 & 30:06:00 & 12.6 & 1\arsec.1/0\arsec.8& 0.179 \\% & 0.105\\
SE3     & 01:36:22.8 & 30:31:30 &  8.2  & 0\arsec.7/0\arsec.6 & 0.184 \\%& 0.107 \\
NW4     & 01:32:37.6 & 31:29:24 & 12.7 & 0\arsec.6/0\arsec.6 & 0.151 \\%& 0.088\\
NW5     & 01:30:28.8 & 31:34:12 & 17.0 & 0\arsec.7/0\arsec.6 & 0.177 \\%& 0.104 \\
NW6     & 01:28:20.0 & 31:39:00 & 22.6 & 0\arsec.7/0\arsec.6 & 0.152 \\%& 0.089\\
NW7     & 01:31:46.9 & 32:20:00 & 25.4 & 0\arsec.6/0\arsec.6 & 0.176 \\%& 0.103\\
BG      & 01:50:50.9 & 30:39:36 &  --  & 0\arsec.6/0\arsec.6 & 0.168 \\%& 0.099\\
\hline \hline
\label{list_clust}
\end{tabular}
\end{center}
\end{table}

\subsection{Photometry}

The images were bias-subtracted, flat-fielded, distortion-corrected, and combined using the standard procedures of the Suprime-Cam Deep Field Reduction  (SDFRED) software \citep{2002AJ....123...66Y,2004ApJ...611..660O}.
The photometry was performed using the \small \textsf{IRAF} \normalsize version of the package \small \textsf{DAOPHOT} \normalsize \citep{1987PASP...99..191S}. A preliminary selection of detections with signal-to-noise ratio greater than 4 was performed with the automatic star-finding algorithm \small \textsf{DAOFIND} \normalsize and the aperture photometry was obtained with the task \small \textsf{PHOT} \normalsize using an aperture radius of 9 pixels for the images taken during the August run (fields SE1 and SE2), which had the poorest seeing (see Table 1), and an aperture of 6 pixels for the others. An empirical point-spread function was built with the task \small \textsf{PSF} \normalsize by selecting several isolated and bright stars in each field. Finally PSF-fitting photometry was extracted with the task \small \textsf{ALLSTAR} \normalsize \citep{1994PASP..106..250S}. To reduce spurious detections, the final catalogs for each field includes only objects with $\chi^2 < 3$, and $\mid\!sharp\!\mid$ $< 0.3$, where $\chi^2$ is a parameter that indicates the quality of the fit, and $sharp$ measures the spatial extent of a detection, being defined as the difference between the square of the width of the measure object and the square of the width of the PSF.

The photometry was calibrated using data for \citet{1992AJ....104..340L} standards observed during the entire run.
Galactic extinction corrections were derived following \citet{1998ApJ...500..525S}.
We applied an average correction corresponding to the extinction at the central position of each field. The adopted values of A$_V$ are displayed in Table 1.

%________________________________________________________________

\subsection{Completeness}

We performed artificial star tests to evaluate the completeness of our photometry. We added artificial stars to the original images
with magnitudes in the ranges 19 mag $< I <$ 26 mag, and 19 mag $< V <$ 27 mag,
%and colours populating the entire CMD,
with a binning step of 0.2 mag that was increased to 0.5 mag for magnitudes brighter than I = 23 mag. A total of 56000 stars were injected at random positions onto each image for both filters. Photometry was then performed in the same way as described in the previous section, and the same selection criteria on the \textsf{DAOPHOT}  parameters were applied.
The final detections were considered as stars if their position was within two pixels of the original list of injected sources, and their magnitude within 0.3 mag of the assigned value. Our photometric data reach the 50\% completeness level at $V \sim 25.6$ mag and $I \sim 24.7$ mag in all fields.

 %________________________________________________________________

\section{Colour-magnitude diagrams}

\subsection{The southeastern fields in the region of the \hi complex}

  \begin{figure}
\includegraphics[width=9cm,bb=25 5 415 415]{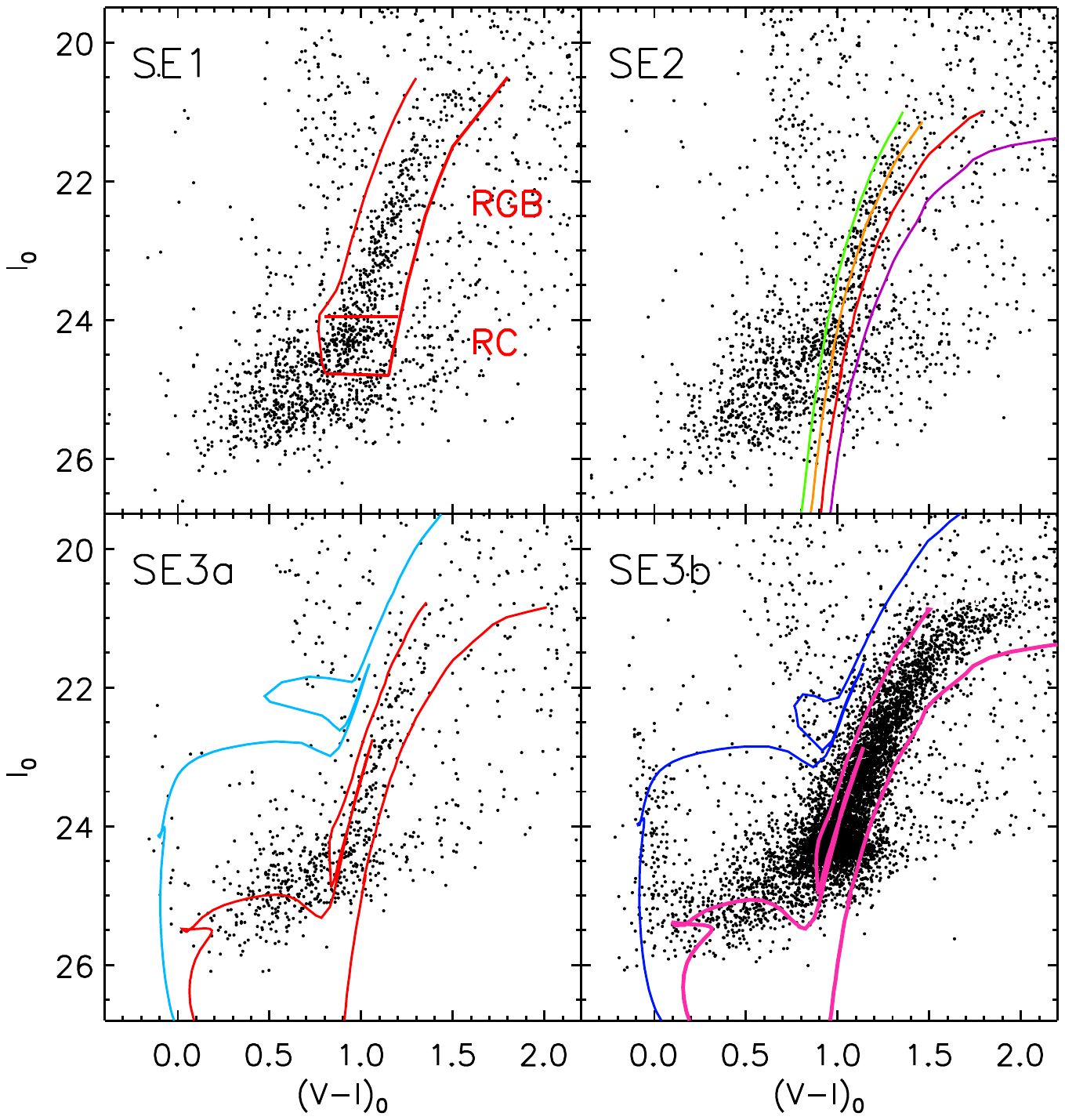}
   \caption{
   CMDs of fields SE1 ({\em top-left}), SE2 ({\em top-right}), SE3a ({\em bottom-right}), and SE3b ({\em bottom-right}).
   The boxes in the {\em top-left} panel display the location of RGB and RC stars in the diagrams. Ten Gyr isochrones from \citet{2008A&A...482..883M}  at a metallicity of \Z = -0.4, -0.7, -1, -1.3 %0.008, 0.004, 0.002, 0.001
   are overlaid on the {\em top-right} panel to give an indication of the expected metallicity range of the RGB population.  Light blue and red tracks in field SE3a ({\em bottom-left}) correspond to a metallicity \Z = -0.7 and an age of 250 Myr, 1 Gyr, and 10 Gyr, while dark blue and purple tracks in field SE3b ({\em bottom-right}) indicate the same age range for a metallicity of \Z = -0.4.
   }
              \label{cmdsouth}%
    \end{figure}

Figure \ref{cmdsouth}
shows the CMD of the three fields in the southeastern region of the galaxy where the main \hi complex  is located (see Fig. 1). The CMDs of field SE1 and SE2, which cover the two main substructures of the southeastern gaseous complex, are displayed in the {\em top-left} and {\em top-right} panels, respectively. Field SE3 maps the region between the \hi feature and the optical disc of M33; the CMD of this area is shown in the bottom panels of the figure.
To differentiate the stars at the edge of the optical disc from a
more diffuse stellar component, we divided the field into two halves (SE3a, and SE3b; see Fig. 1) and display the corresponding diagrams in the {\em bottom-left} and {\em bottom-right} panel of the figure, respectively.

The most prominent feature is the red giant branch (RGB), corresponding to stars with colours $(V-I)_0$ between 0.7 and 2 mag, and $I_0 < 24$ (see {\em top-left} panel in Fig. \ref{cmdsouth}). The RGB is clearly detected in all fields, although in SE2 and SE3a this feature is relatively less populated, especially in the upper end.
The edge of the optical disc of the galaxy is shown in panel SE3b as indicated by the redder and more populated RGB.

Because of the well-known age-metallicity degeneracy, different combinations of age and metal abundances are consistent with the colour of this feature.
This is shown in the {\em top-right} panel where
the RGB of SE2  is compared to isochrones from \citet{2008A&A...482..883M} with a common age of 10 Gyr and metallicities of \Z = -0.4, -0.7, -1, -1.3   %0.008, 0.004, 0.002, 0.001
(from redder to blue colours), where \Z = log(Z/Z$_{\odot}$) and Z$_{\odot}$ = 0.019. %to show the range of metal abundances compatible with its colour.
On the other hand, the width of the RGB could also be due to an age spread as the two isochrones at 1 and 10 Gyr (with \Z = -0.7 in SE3a, and  \Z=-0.4 in S3b) show. In particular for SE3b, the age range of the stellar populations may be several Gyrs.

   \begin{figure}
\includegraphics[width=9cm,bb=25 5 415 415]{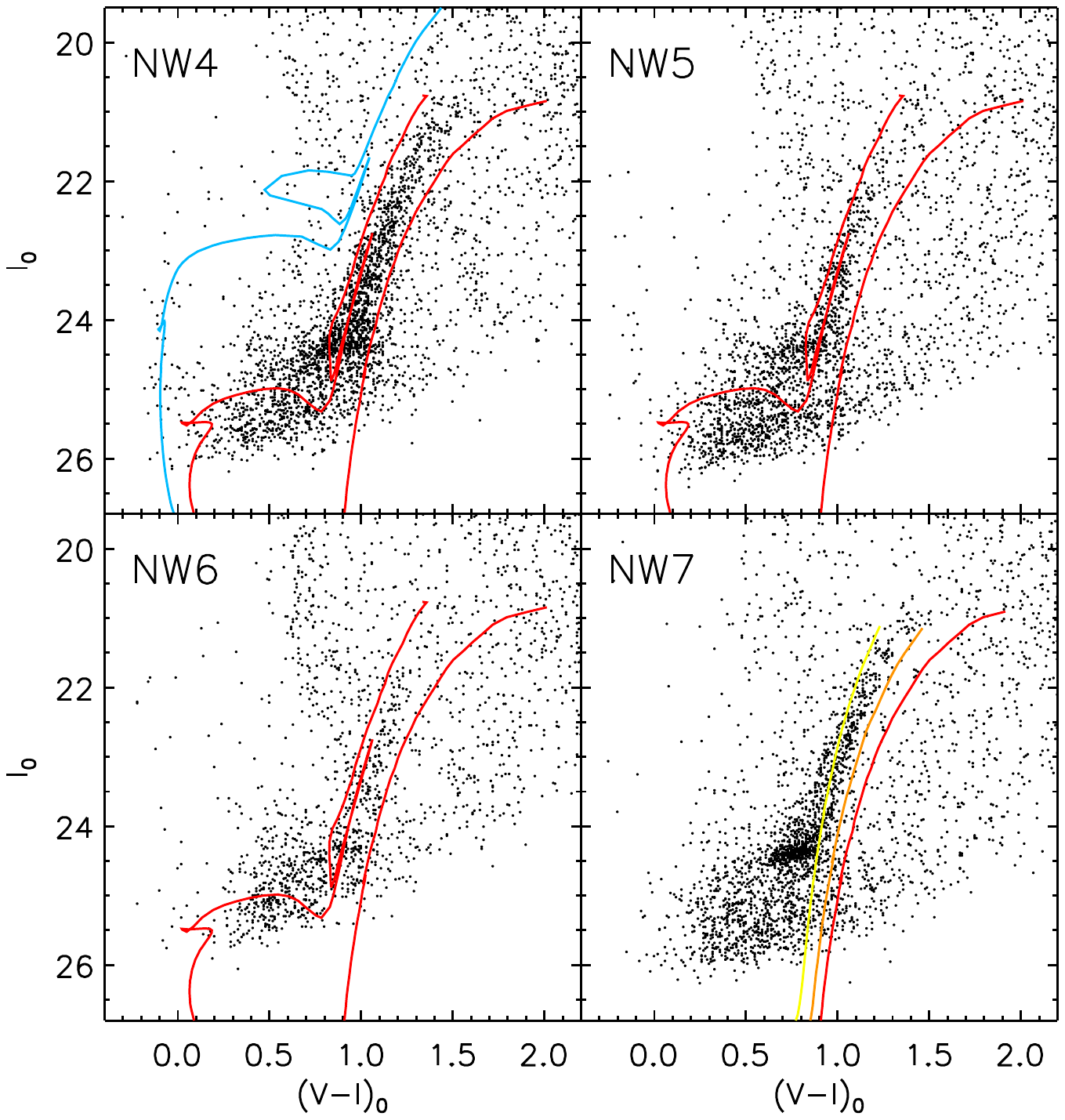}
   \caption{CMDs of field NW4 ({\em top-left}), NW5 ({\em top-right}), NW6 ({\em bottom-left}), and NW7 ({\em bottom-right}). Isochrones from \citet{2008A&A...482..883M} are overlaid on the diagrams. Blue and red tracks in the first three panels indicate a metallicity of \Z = -0.7 and ages of 250 Myr, 1 Gyr, and 10 Gyr. The same metallicity and age range is shown in the diagrams to mark the change in the RGB shape from NW4 to NW6. Isochrones in field NW7 ({\em bottom-right}) correspond to an age of 10 Gyr and a metallicity of \Z = -0.7, -1, -1.7 (from right to left) and give an indication of the expected metallicity range in the RGB stellar population.}
              \label{cmdnorth}%
    \end{figure}

   \begin{figure}
   \begin{center}
\includegraphics[width=7cm,bb=30 8 410 410]{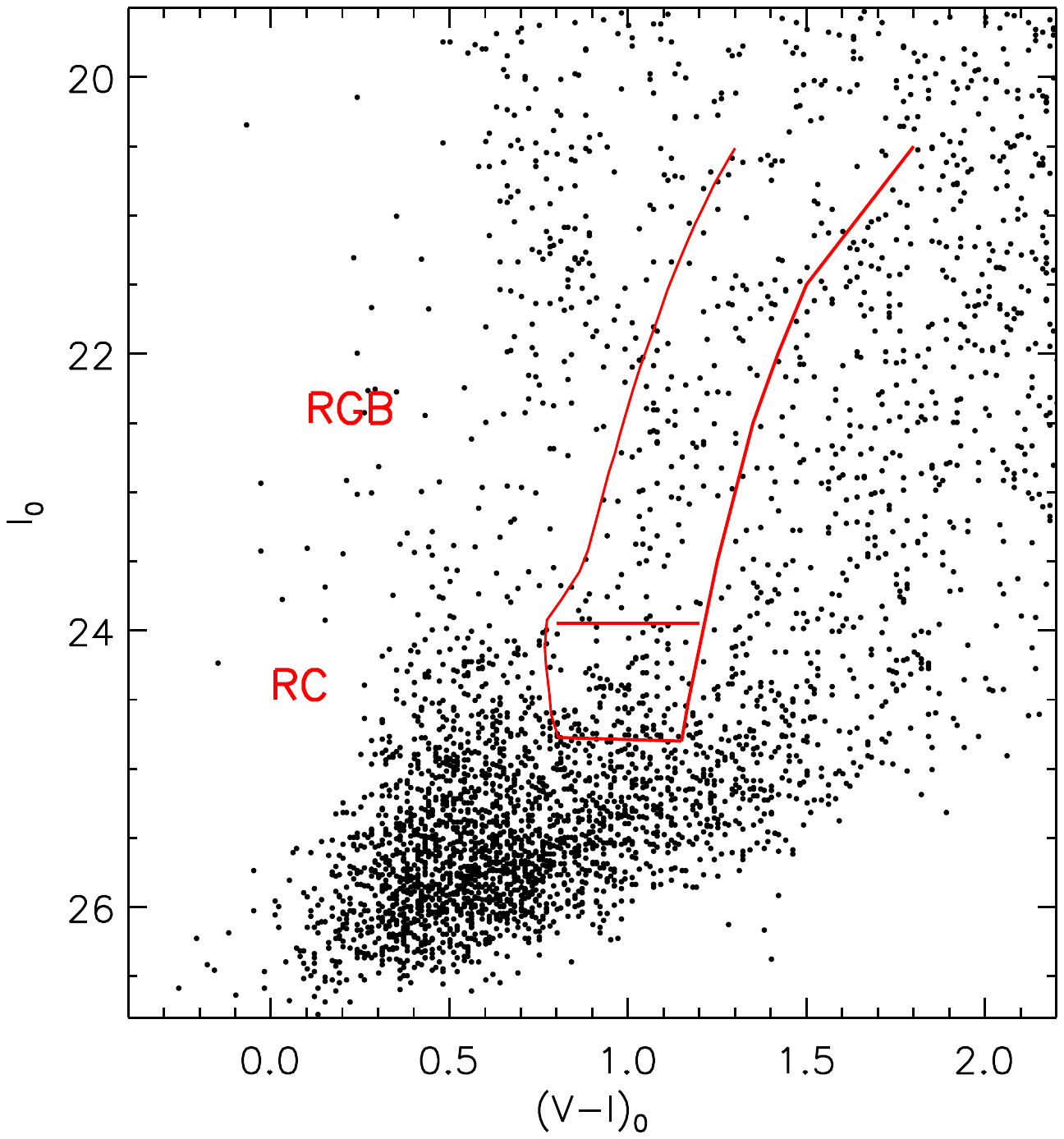}
   \caption{CMD of the control field used to inspect the foreground and background contamination.}
              \label{cmdbg}%
    \end{center}
    \end{figure}

Figure \ref{cmdsouth} also indicates a population of stars in the red clump (RC) phase, i.e. intermediate-age  (1-10 Gyr) stars experiencing a core-helium burning phase. The average magnitude of this feature is around $I_0 \sim 24.4$ mag, and the colour range is $0.7<(V-I)_0<1.2$ (as shown by the box in the {\em top-left} panel of Fig. \ref{cmdsouth}).
In addition the RC mean colour and magnitude depend on the age and metallicity of the stars \citep{2001MNRAS.323..109G}, although the magnitude of the RC is more sensitive to age rather than to metallicity \citep{2010ApJ...718.1118D}, and its colour, for stars older than 2 Gyr, is mainly sensitive to metallicity variations \citep{2001MNRAS.323..109G}. We use these properties of the RC feature to constrain the age and metallicity of these stars in the different fields (see Sect. 5).

The CMDs of SE3 show a small population of main-sequence blue stars with ($V-I$)$_0$ $\lesssim$ 0 that are both related to the edge of the optical disc (SE3b), and scattered throughout the field out to a maximum projected distance of 50\arminsp (SE3a), %16 kpc deprojected
about ten times the disc scale \citep[6\armin.4 in the $J$ band;][]{1994ApJ...434..536R}.
Comparing the brightest magnitude of the blue stars (i.e. the main-sequence turn-off luminosity) to isochrones with \Z=-0.7 implies an age around 250 Myr.

%________________________________________________________________

\subsection{The northwestern fields}

\subsubsection{Stellar populations in the \hi warp}

Fields NW4-NW6 map the northwestern side of the galaxy along the \hi warp out to the northernmost \hi clouds found around M33.
The CMDs of field NW4 (Fig. \ref{cmdnorth} {\em top-left} panel) show a stellar population with similar properties as the southern regions at a comparable distance from the centre of M33 (SE2). On the other hand, the RGB of NW5 and NW6 (Fig. \ref{cmdnorth}, {\em top-right} and {\em bottom-left} panels) gradually becomes bluer, narrower, and less populated, implying a smaller age spread in the RGB, and/or a change in the average metal abundance. Isochrones at 1 and 10 Gyr with \Z = -0.7 are overlaid as a comparison.
A small population of blue main-sequence stars is also detected in NW4 and NW5, out to a maximum angular distance of 60\arminsp. % ($\sim$ 15 kpc).

   \begin{figure}
\includegraphics[width=7.5cm,bb=25 5 415 415]{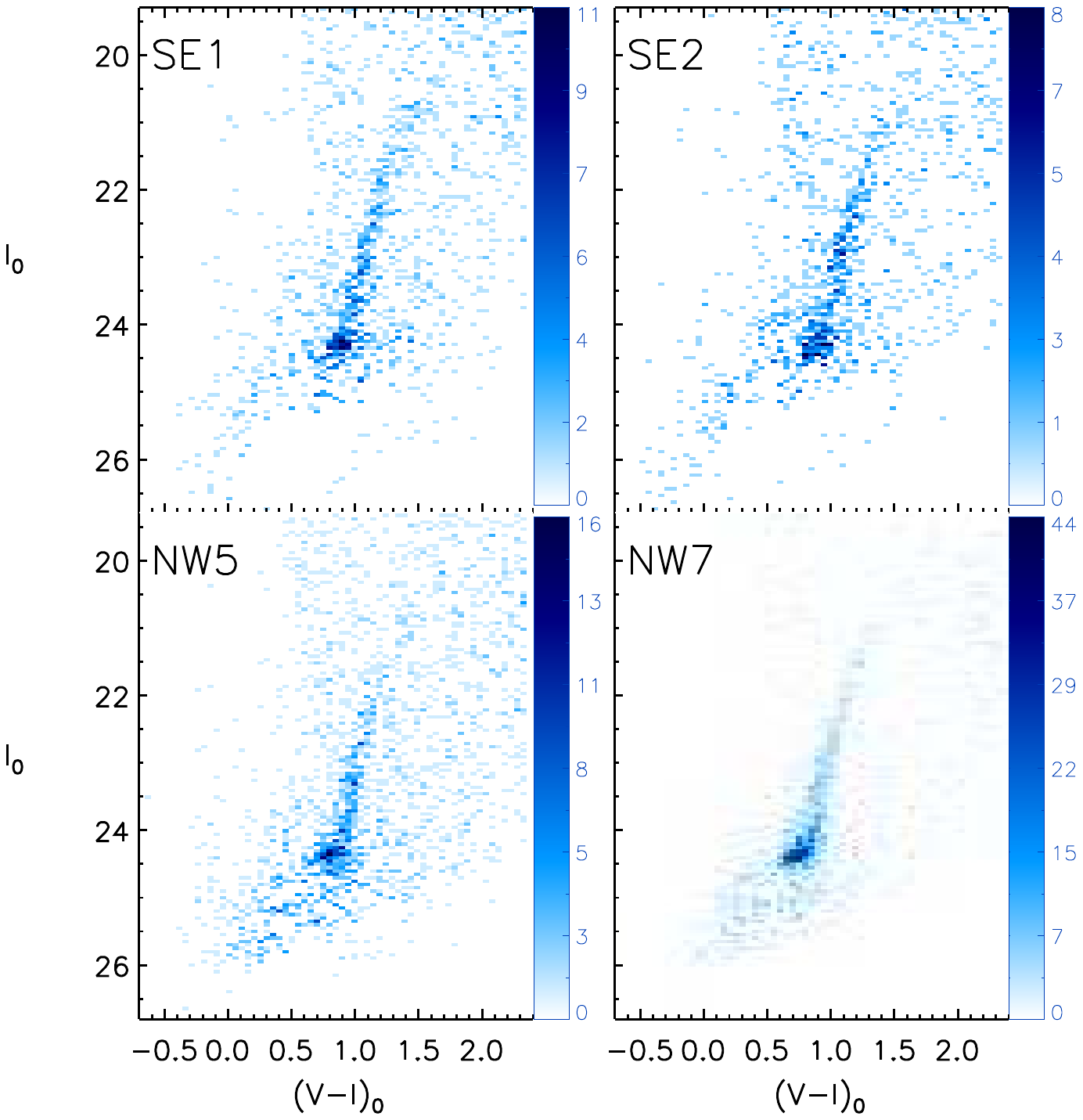}
   \caption{Contamination-subtracted Hess diagrams of field SE1 ({\em top-left}), SE2 ({\em top-right}), NW5 ({\em bottom-left}), and NW7 ({\em bottom-right}). The resolution of the diagrams is 0.05 $\times$ 0.05 mag, and the colour scale indicates the number of stars per pixel. Only positive residuals are shown in the diagrams.}
              \label{hess}%
    \end{figure}

 \begin{figure*}
   \centering
 \includegraphics[width=11cm,bb=120 0 550 340]{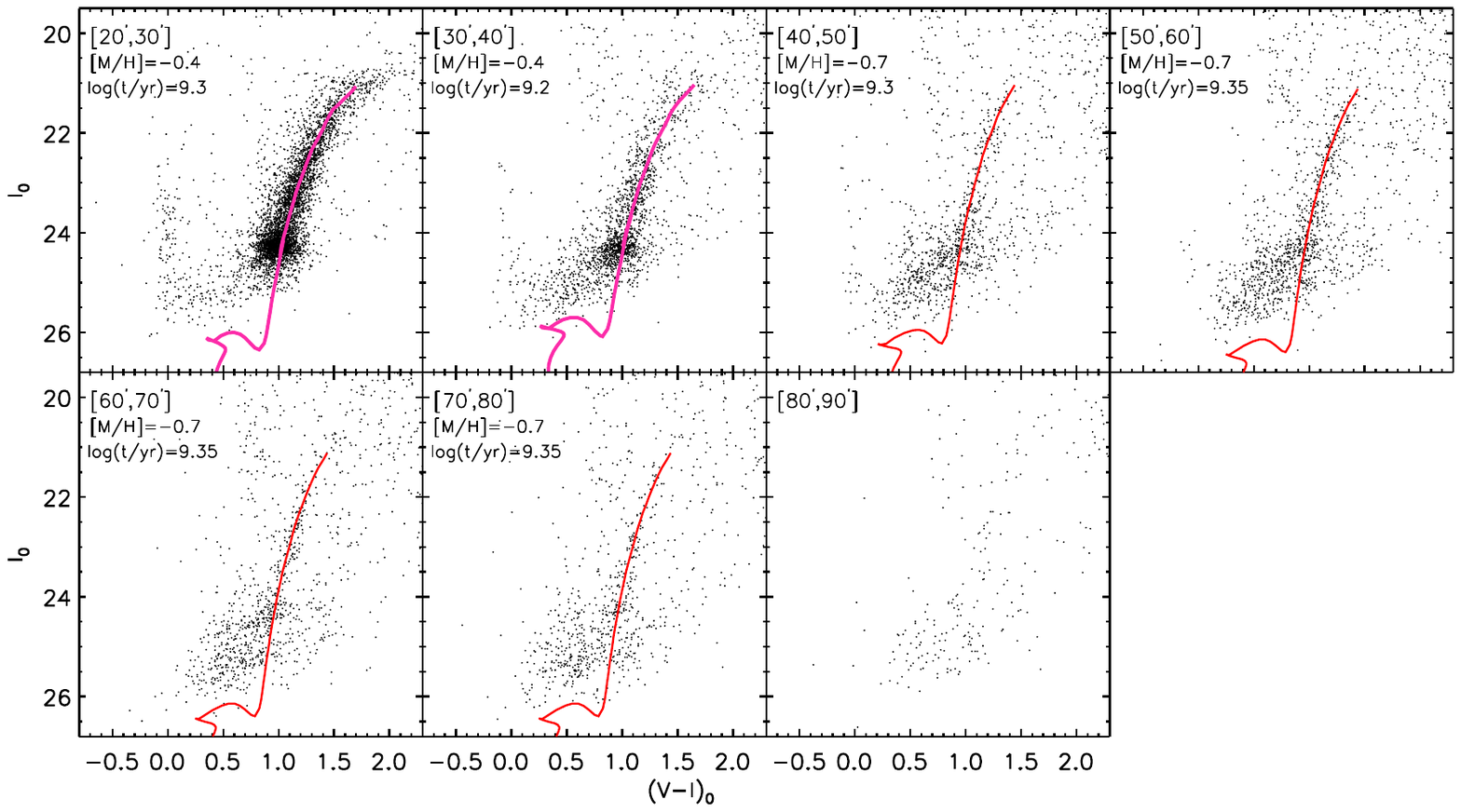} %fig grande
   \caption{CMDs of southern fields in radial bins of width 10\armin. The radial angular distance from the M33 centre is displayed at the top-left corner of each plot. The isochrone that best fits the RGB shape and colour is also shown. The corresponding metallicity and age are indicated in the top-left corner of each panel.}  \label{rad_cmd_south}%
    \end{figure*}

\subsubsection{Stellar populations in the large stellar feature around M33: the NW7 field.}

Field NW7 is not related to any \hi feature
since this region was not included in our 21-cm survey of M33,
but was selected to investigate the stellar population of the puzzling large stellar feature around M33 found by the PAndAS survey. It is roughly one degree further north in declination than NW6. Despite being at approximately the same angular distance from the centre of M33,
the difference between the CMD of these two fields is striking. NW7 shows a clearly detected RGB, and RC, in contrast to NW6, where the RC is hardly visible.
While the number of RGB stars decreases from NW4 to NW6, NW7 shows a peculiar increase in the stellar density,
given that the corresponding projected radius is around 25 kpc (for a M33 distance of 840 kpc).
Moreover,
the overall colour of the RGB and RC features is bluer than in the other fields as can be inferred from the comparison to the stellar isochrones. Population synthesis models with a common age of 10 Gyr and \Z = -0.7, -1, -1.7 are overlaid on the CMD in the {\em bottom-right} panel of Fig. \ref{cmdnorth}. They show that the stellar metal abundances could range between \Z=-1.7 and  \Z=-1,
suggesting the presence of a population with different properties from the other fields.

%________________________________________________________________

\subsection{Foreground and background contamination}

We observed a field offset
by 4 degrees from the centre of M33 to take into account the foreground and background contamination due to Milky Way stars and unresolved distant galaxies misclassified as point sources. The field was processed in the same way as the others and the same cuts were applied to the final list of detections.  The CMD of this region is displayed in Fig. \ref{cmdbg}, and shows that contamination mainly occurs in the region between $0.3 < (V-I)_0 <0.8$ and $I_0 > 24$ mag, and at colours redder than $(V-I)_0 \sim 1.5$ mag. Foreground and background contaminants may affect the lower part of the RC, %as we will discuss later,
while the region of the RGB and blue main sequence are not severely contaminated. At \vi $< 1 $ mag and  $I_0 > 23$ mag, the main sequence of stars in the Galactic halo is clearly visible, but mainly contaminates the blue and bright region of the CMDs, where we do not find significant stellar populations related to M33.
To assess the contribution from foreground and background contaminants, we built the Hess diagram of the control field, providing the density of stars at different positions in the CMD, dividing the diagram into bins of dimension 0.05 $\times 0.05$ mag in colour and $I$ magnitude. We then subtracted it from the Hess diagrams of the other fields around M33, and, as an example, in Fig. \ref{hess} we show the results for four regions. The diagrams reveal that both the RGB and RC can be clearly discerned after the subtraction, confirming that these features, and particularly the RC, are reliable detections.

%________________________________________________________________

 \begin{figure*}
   \centering
    \includegraphics[width=11cm,bb=120 0 550 340]{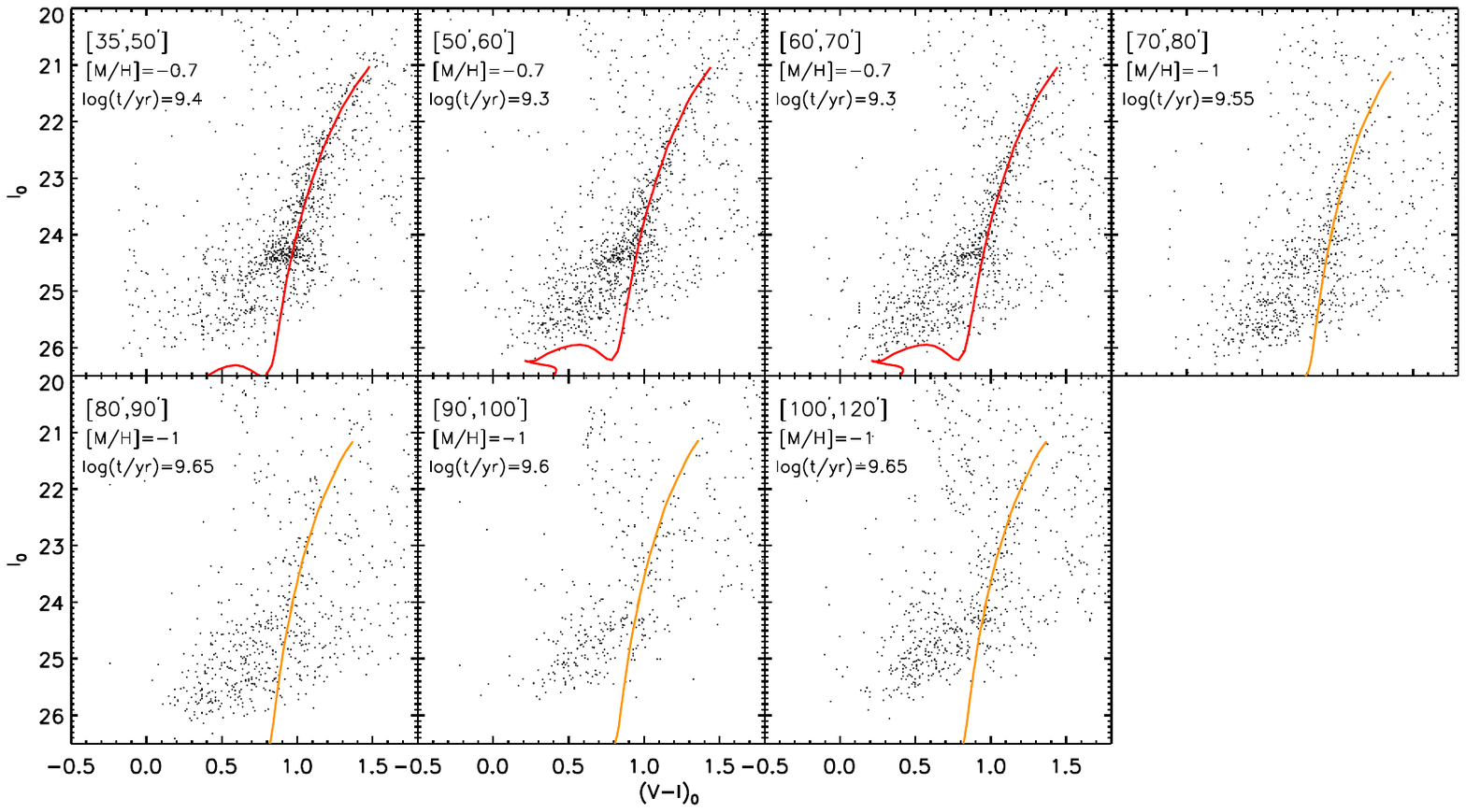}   %fig per astro-ph
   \includegraphics[width=11cm,bb=120 0 550 180]{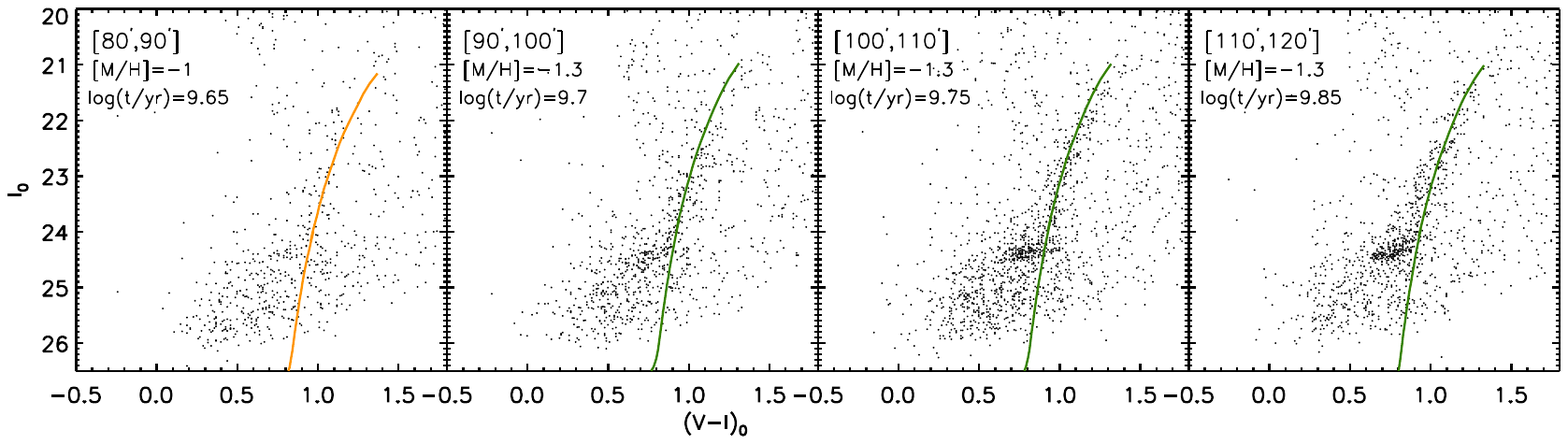}   %fig per astro-ph
   \caption{{\em Top:} CMDs of fields NW4, NW5, and NW6 in radial bins of width 10\armin. The radial angular distance from the M33 centre is displayed in the top-left corner of each plot. The isochrone that best fits the RGB slope and colour is also shown. The corresponding metallicity and age are indicated in the top-left corner of each panel. {\em Bottom:} CMDs in radial bins at angular distances greater than 80$^{\prime}$ including stars in field NW7. The stellar populations in these regions appear to be older and more metal-poor with an average metallicity [M/H]= -1.3.}  \label{rad_cmd_north}%
    \end{figure*}

\section{The distance to M33 from the tip of the red giant branch}

Estimates of the distance to M33 obtained with different methods vary considerably in the literature, ranging between 800 and 940 kpc \citep{2002ApJ...565..959L,2004ApJ...614..167C,2004MNRAS.350..243M,2004AJ....128..224T,2006AJ....132.1361S}.
Here we derive the distance to M33 using the tip of the red giant branch (TRGB) as a standard candle
 %used to measure distances to galaxies of any morphological type
\citep{1993ApJ...417..553L}.
The condition for the correct application of this technique is that
the observed RGB should be well populated, with more than $\sim$ 100 stars within 1 mag of the tip
\citep{1995AJ....109.1645M,2001ApJ...556..635B}.
Thus we built the $I$ luminosity function of the RGB in the CMD of field SE3b, which is the one that most closely fulfils this criterion, for all the stars redder than $(V-I)_0$ $> 0.7$.
We then applied an edge detection Sobel filter to the luminosity function to locate the position of the tip.
The peak of the filter response is taken as the best estimate of the TRGB magnitude, while the half width
at half maximum of the same peak ($\pm$ 0.08 mag) and the photometric error at this magnitude ($\pm$ 0.03 mag) are taken as the associated uncertainties.
An additional source of error is given by the uncertainty in the zero-point coefficients obtained from the photometric calibration, which in the  $I$ band is 0.05 mag.
The tip is found at $I_0$ = 20.73 $\pm$ 0.1, with all errors combined in quadrature. Assuming an absolute magnitude of the TRGB of M$_I^{TRGB}$ = -4.02 $\pm$ 0.05 \citep{2004AJ....128..224T}, this corresponds to a distance modulus $(m-M)_0 = 24.75 \pm$ 0.11, implying a distance to M33 of 891 $\pm$  45 kpc.
However, field SE3b contains stars in the farthest side of the optical disc of M33.
As one can see from Fig. 6, the majority of
    the RGB stars used to determine the tip reside between 20 and 30 arcmin, corresponding to a projected distance of ~ 6 kpc
    or a deprojected radius of 10 kpc (being the field close to the minor axis). Thus the stars are about 8 kpc more distant
    than the center of M33, hence the implied distance to M33 is 883 $\pm$ 45 kpc. The uncertanties in the deprojection give an
    additional source of error of 2 kpc, which once added in quadrature, do not significantly increase the uncertainty in the distance.

This  distance modulus agrees with the results of ground-based observations of \citet{2004AJ....128..224T}, $(m-M)_0$ = 24.69 $\pm$ 0.07, with HST observations of M33 disc stars, ($m-M$)$_0$ = 24.80 $\pm$ 0.04 (random)$^{+0.15}_{-0.11}$(systematic), \citep{2002AJ....123..244K}, and,
within the errors, is compatible with the distance value of 840 $\pm$ 40 kpc  \citep{2001ApJ...553...47F} that we assumed in Grossi et al. (2008).

%________________________________________________________________

\section{Constraints on the age and metallicity of the stellar populations}

\begin{figure*}
   \centering
 \includegraphics[width=9cm,bb=20 5 595 540]{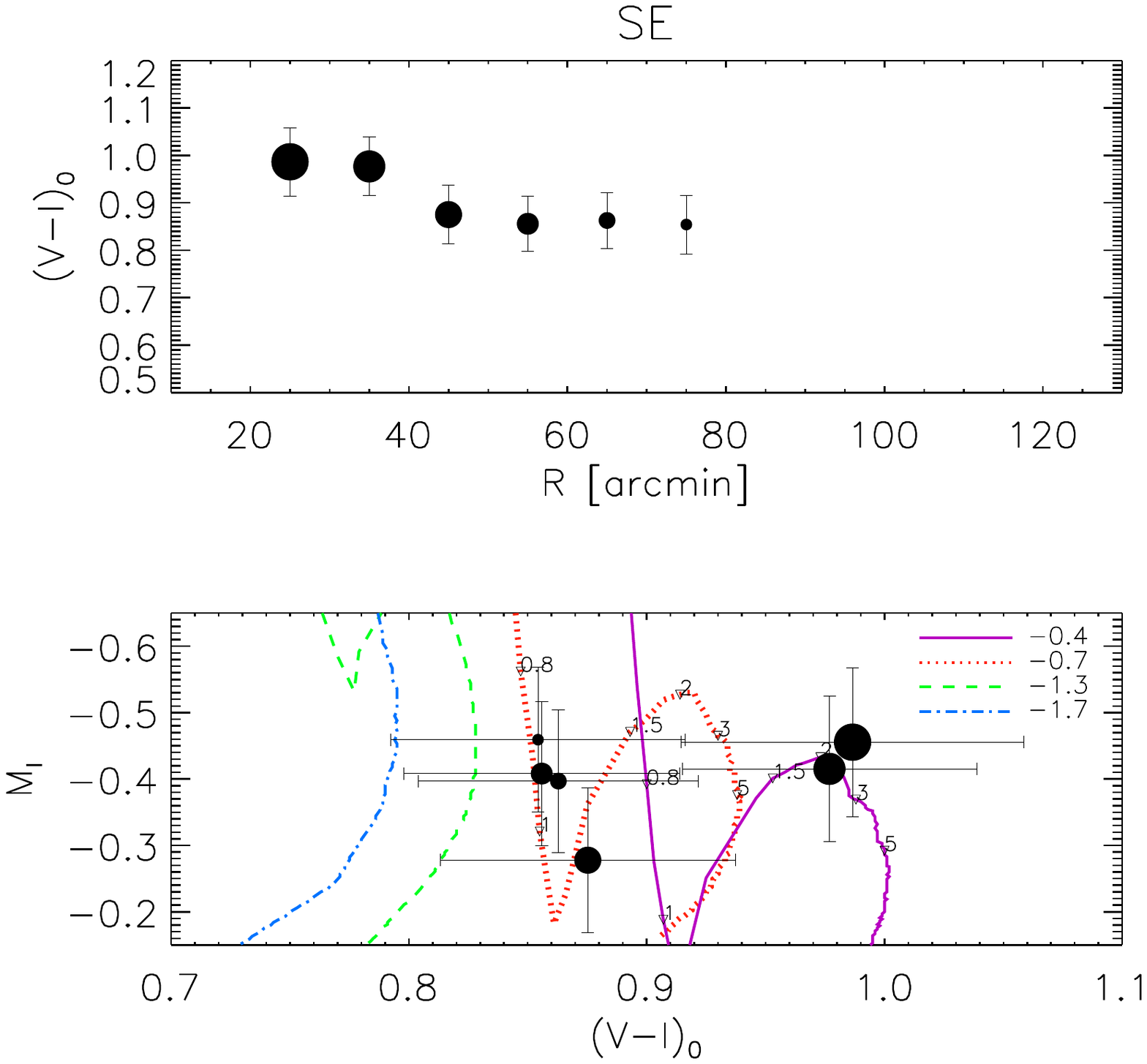} %fig grande
   \includegraphics[width=9cm,bb=20 5 595 540]{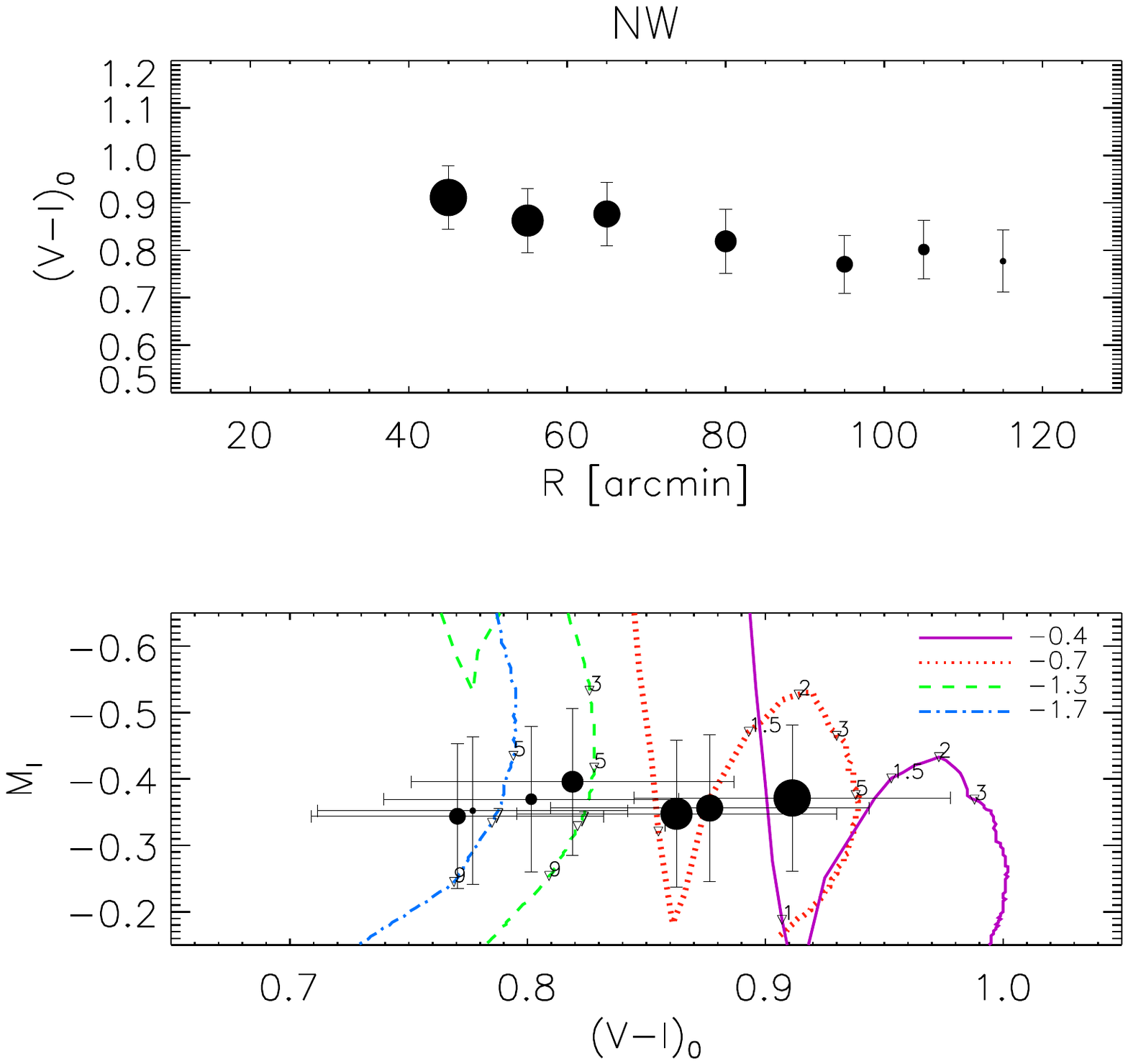}   %fig per astro-ph
   \caption{Mean colour of the RC as a function of the angular distance from the centre of M33 in the southern ({\em top-left}) and northern ({\em top-right}) fields. The size of the dots decreases with the distance to the M33 centre. A clear colour gradient is found in both regions. % fields, while the change in metallicity is more shallow in the southern region.
   Mean theoretical RC colour and magnitude for different metal abundances (from \Z=-0.4 to \Z=-1.7) and ages are displayed in the {\em bottom} panels. The numbers along the model indicate different ages in Gyr units. The size of the filled dots in the lower panels is the same as in the upper panel to identify the different radial bins.
      }  \label{RC_col}%
    \end{figure*}

Evolved RGB and RC stars appear in a stellar population after 0.9 - 1.5 Gyr: stars first enter the RGB phase, where hydrogen is burnt in a shell around the helium core, then move to the RC when the core-helium-burning stage begins. Even though the RGB and RC are affected by the age-metallicity degeneracy, we can use them together to constrain the mean metallicity and age of the stellar populations. As mentioned in Sect. 3.1, any variations in the RGB colour and slope are more sensitive to metallicity; the magnitude of the RC has a stronger dependence on age \citep{2010ApJ...718.1118D}, while the colour of the RC for ages older than 2 Gyr is more sensitive to metallicity variations \citep{2001MNRAS.323..109G}. Bearing this in mind, we can use these features to perform a qualitative inspection of the CMDs at different radii to estimate the mean properties of the stellar populations as we proceed to larger galactocentric distances. Thus, we built CMDs in radial bins of 10 arcmin width and compared the colour, the magnitude, and the shape of these features to synthetic isochrones and models %from \citet{2008A&A...482..883M}, and \citet{2001MNRAS.323..109G}
to investigate possible radial gradients in age and metallicity.

\subsection{Age and metallicity gradients from the red giant branch}

Figures \ref{rad_cmd_south} and \ref{rad_cmd_north} show the spatial variation in the CMDs for the southern and northern fields, respectively. In each diagram, we compared the RGB feature within the magnitude range 20.7 mag $< I_0 < 24$ mag to a set of stellar isochrones with metallicity between \Z =-1.7 and \Z=-0.4 \citep{2008A&A...482..883M}.
For each set of isochrones, we calculated the colour difference $\delta(V-I)$ between the observed RGB and the set of fiducial stellar tracks. The track that minimises the rms of the distribution of $\delta(V-I)$ is selected as an indicator of the mean age and metallicity of the stellar population in a radial bin.

The average metallicity in the SE1-SE3 fields (Fig. \ref{rad_cmd_south}) changes from \Z=-0.4 at the edge of the optical disc ($R <$ 40$^{\prime}$, first two panels), to \Z=-0.7 at angular distances  $R >$ 40$^{\prime}$, and the same abundance is obtained out to the outermost bin where a weak RGB is detected ($70^{\prime} < R < 80^{\prime}$). The corresponding mean age of the populations at large radii ( $R >$ 40$^{\prime}$) does not vary considerably with the distance, being around log(age)=9.3-9.35.  % and 9.45 with an uncertainty of $\pm 0.05$.
A blue plume of main-sequence stars is detected out to the third radial bin $40^{\prime} < R < 50^{\prime}$.

The properties of the stellar populations in the NW fields (Fig. \ref{rad_cmd_north}) appear similar to the SE ones when one considers the same range of angular radii ($35^{\prime} <R < 70^{\prime}$): the average metallicity is found at \Z=-0.7 and the stellar ages in these radial bins vary between 2 and 2.5 Gyr (9.3 and 9.4 in logarithmic units). Young main-sequence stars are also detected in these fields out to a radial distance of 60$^{\prime}$.
The outer radial bins instead, at $R > 70^{\prime}$, show an older and more metal-poor RGB stellar population, evidence that becomes more compelling when stars from NW7 are included (Fig. \ref{rad_cmd_north} {\em lower panel}).
The lower metallicity of the RGB stars at these radii is suggested by the overall bluer colour than the stars in the bins closer to the M33 centre. While the RGB and RC features become weaker from NW4 to NW6 ({\em upper panel} of Fig. \ref{rad_cmd_north}),
%The figure shows that, contrary to what is found in NW6, where the RGB becomes weaker at large radii,
there is a clear increase in the number of stars in NW7 out to the last radial bin ($110^{\prime} < R < 120^{\prime}$).
The isochrones that most closely reproduce the slope and colour of the RGB have \Z=-1 in the NW6 field (Fig. \ref{rad_cmd_north}, {\em upper panel}), whereas when stars in NW7 are included ({\em lower panel}), the average population at $R > 90^{\prime}$ appears to be even more metal-poor (\Z = -1.3) and older (between 5 and 7 Gyr).

\subsection{Age and metallicity gradients from the red clump}

We derived the mean photometric properties of the RC as a function of radius following \citet{2001MNRAS.323..109G}.
We determined the histogram of the $I$  magnitude and colour distribution of stars in regions of the CMDs including the locus of the RC, located within 24 mag $ < I_0 < 24.7$ mag and $0.7$ mag $ < (V-I)_0 < 1.2$.
We then fitted a Gaussian plus a quadratic background function to both histograms, taking the central values of the Gaussians as the mean magnitude and colour of the clump.
Given the difficulty in identifying a clear clumping of stars in the RC region in
the third and fourth bin of the SE fields, or the
low number of RC stars in the radial range between 70\arminsp and 90\arminsp in the NW ones,
the results of our analysis at these radii should be treated with caution.
To maximize the number of stars in the RC area, the range of angular distances between
70\arminsp and 90\arminsp  was considered as one single bin, and
the outermost regions ($R > $ 90\armin) include detections from both NW6 and NW7.

The results of our analysis are shown in Fig. \ref{RC_col}.
In the {\em upper} panels, we plot the RC colour versus radius for the SE ({\em top-left}), and NW ({\em top-right}) fields. The size of the dots decreases with the distance to the M33 centre to help us to identify the
radial bins in the lower panel plots.
In both regions, there is a clear change in $(V-I)_0$ with radius. %, which is mainly related to a variation in metallicity.
In the {\em lower} panels, the mean RC colour and magnitude are compared to the models of \citet{2001MNRAS.323..109G} to assess more reliably the range of age and abundances of the populations.
The models show that variation in the metal abundances is the main driver for the colour change: the metallicity ranges from \Z=-0.4, \Z=-0.7 in the inner bins, down to \Z=-1.3/-1.7 in the outermost ones. This agrees with what we found from the RGB analysis. However, the error in the photometry (and in the distance modulus for $M_I$) displayed in the figure %the half bin width in deriving the luminosity and colour function of 0.025 mag and
make the determination of the age subject to some uncertainties.
%but also the age of the stellar population changes with radius.
In the SE fields, the redder bins correspond to stars with ages around 2 Gyr and \Z=-0.4, while at 40\arminsp $< R <$ 80\arminsp the colours and mean magnitude of the RC are compatible with an age between 0.8 Gyr and 1.5 Gyr, slightly lower than what we found for the RGB stars, and a metal abundance of \Z=-0.7. In the NW fields, the colour and magnitude of the RC in the first bin may correspond to a population with either a mean age of $\sim$ 2 Gyr and \Z=-0.7, or an age between 1 Gyr and 2 Gyr with a higher abundance (\Z=-0.4). At 50\armin $< R <$ 70\armin, a metallicity of \Z=-0.7 would imply an age between 1 and 2 Gyr, although even older ages would be compatible with the photometric errors.

At larger radii ($R >$ 70\armin) the NW data indicate that the properties of the stellar population clearly differ. The colour becomes bluer than $(V-I)_0 = 0.85$, corresponding to ages older than 5 Gyr and an abundance of \Z=-1.3 (70\armin $< R <$ 90\armin) or lower in the most distant bins at $R >$ 90\arminsp (\Z=-1.7).
In the following section, we try to more tightly constrain the range of ages and metallicities of stars in the outermost field by comparing the observed CMDs to models with a range of star formation histories (SFHs).

%________________________________________________________________

\begin{figure}
   \centering
   \includegraphics[width=8cm,bb=0 8 430 215]{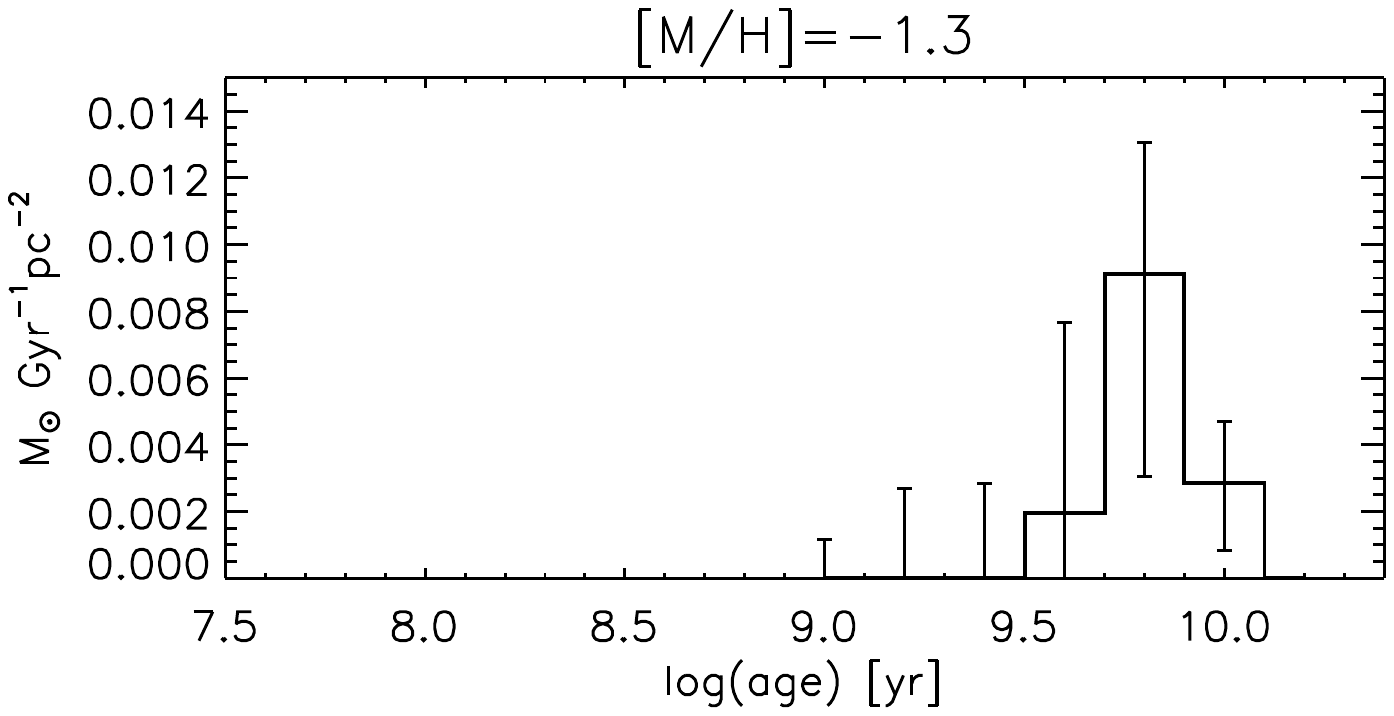}   %fig per astro-ph
   \includegraphics[width=9cm,bb=30 0 415 280]{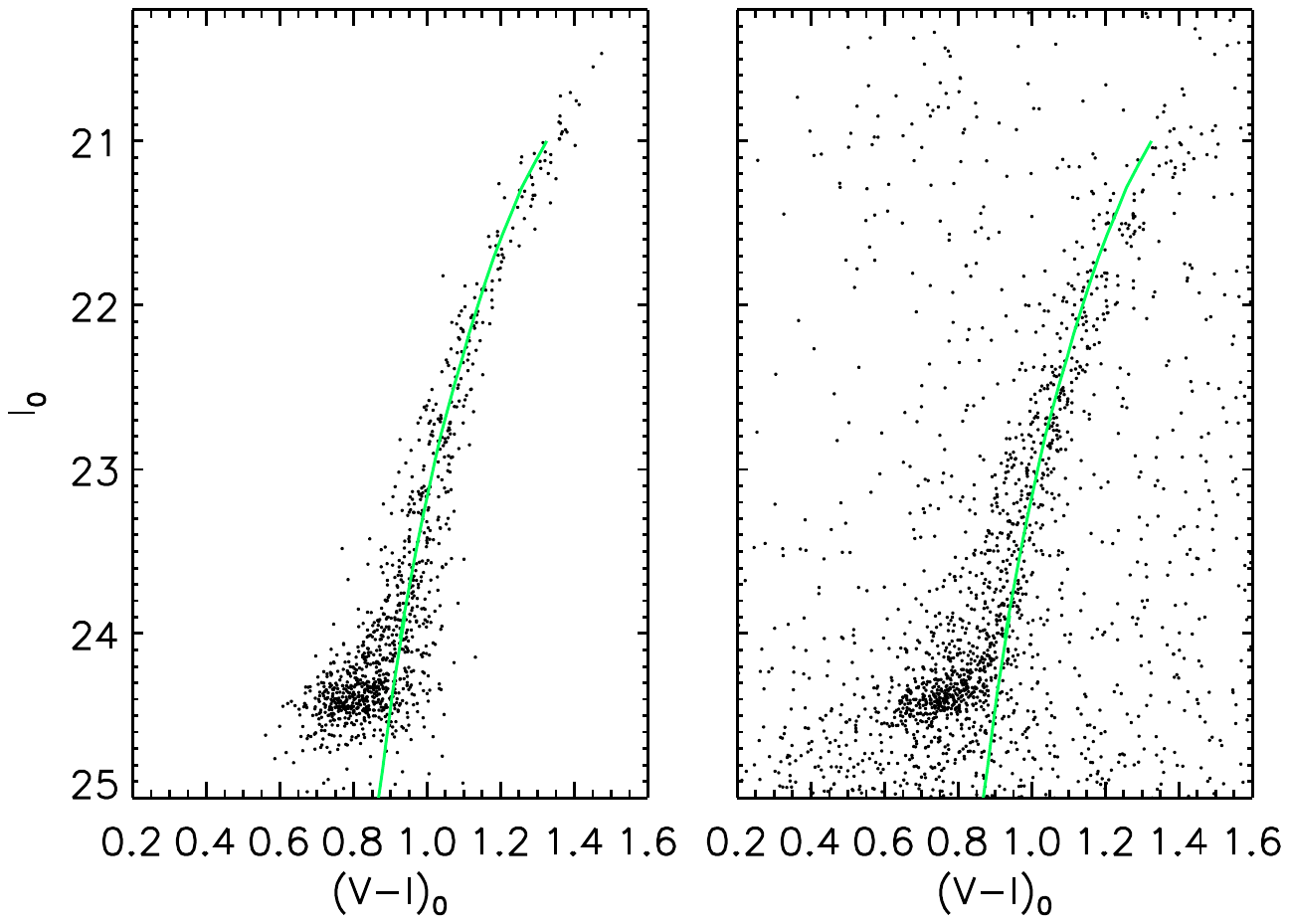}   %fig per astro-ph
\caption{Star formation rate as a function of time of field NW7 obtained from the CMD fitting analysis ({\em top}), and the resulting synthetic CMD ({\em bottom-left}) compared to the observed one ({\em bottom-right}). A 6.3 Gyr (log(age) = 9.8) isochrone with \Z = -1.3 is overlaid on the RGB of both diagrams.
   }  \label{sfh}%
    \end{figure}

\begin{figure}[h]
   \centering
 \includegraphics[width=8.5cm, bb= 50 0 420 480]{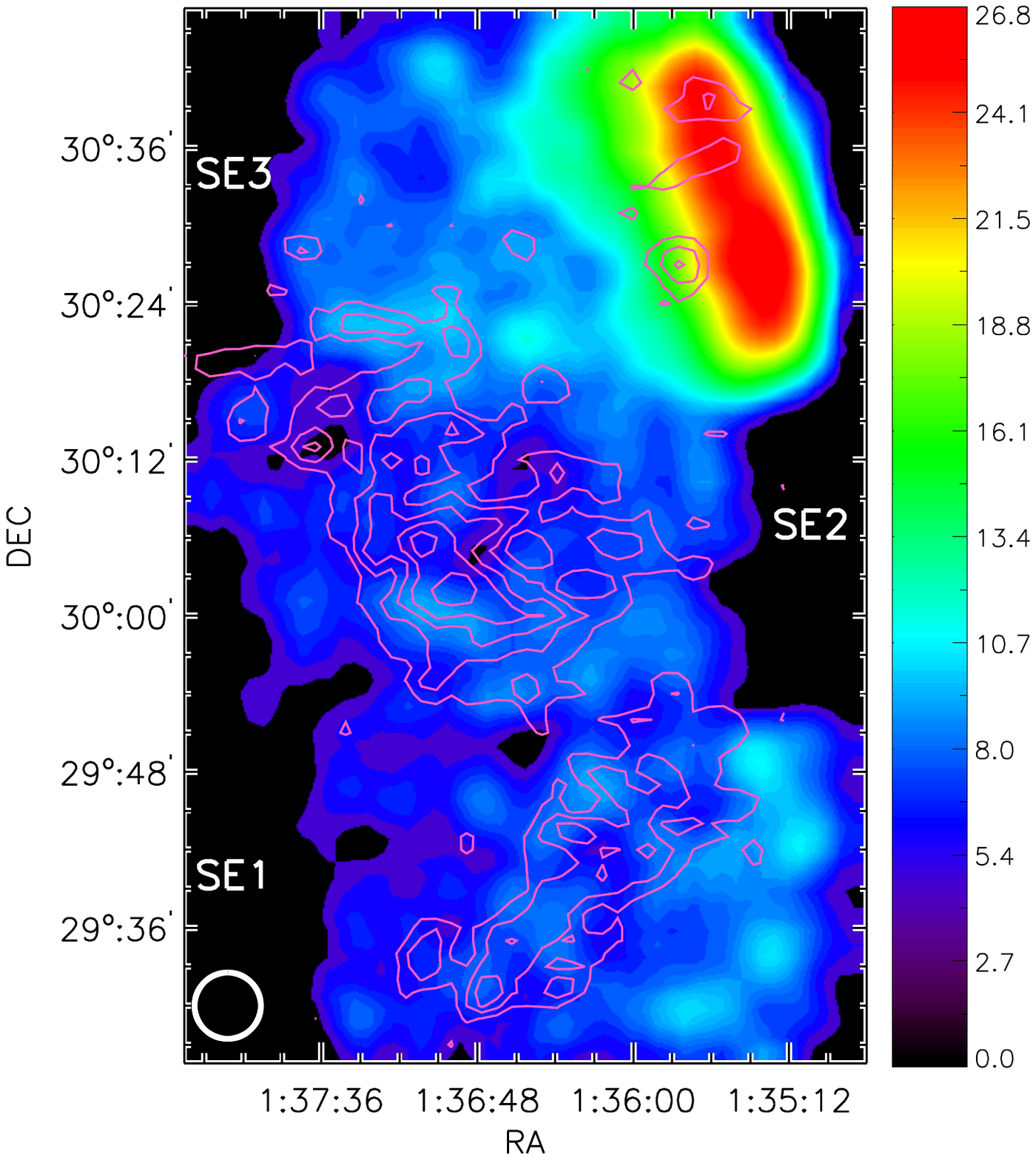} %fig grande
   \caption{Spatial distribution of RGB and RC stars with 20.7 mag $< I <$ 24.6 mag in the southern fields.
   The resolution of the map is 1\armin $\times$ 1\armin, smoothed with a gaussian kernel of FWHM five pixels (shown at the bottom-left corner of the figure). The colour scale indicates the number of stars per pixel. The stellar distribution is compared to the \hi complexes detected in this region. Contours are drawn at 3 to 10.5 $\times 10^{18}$ cm$^{-2}$ in steps of 1.5 $\times 10^{18}$ cm$^{-2}$.
   }  \label{dens_map_south}%
    \end{figure}

\section{Star formation history of field NW7}

A more accurate constraint of the properties of the stellar populations in the northernmost region may help us to untangle the origin of the stellar feature around M33. % particularly as regards a possible interaction scenario
Therefore, we applied the method of synthetic CMD fitting to the diagram of NW7 to derive the best-fit SFH of the stellar populations in this region.  %, by comparing the data to linear combinations of model CMDs with a defined range of age and metallicities.
%with M31.

We used the StarFISH package \citep{2001ApJS..136...25H}, which determines the best-fit CMD according to a maximum likelihood statistic.
To build the model CMDs, we selected four sets of isochrones with metallicities \Z=-1.7, -1.3, -1.0, -0.7, and ages between 1 and 10 Gyr (with a time binning step of log(age) = 0.2), as derived from the analysis of the RGB and RC features in Sect. 5. The code populates the set of stellar isochrones  \citep{2008A&A...482..883M}  assuming a given initial mass function \citep{1955ApJ...121..161S}, shifting the magnitudes according to the distance modulus, and applying photometric scatters derived from the artificial star tests (see Sect. 2.2 for
details).
Each synthetic CMD with a given range of ages and a fixed metallicity can be linearly combined to construct a model diagram  for any arbitrary SFH.
The code then determines the best-fit SFH performing $\chi^2$ minimization of the differences in
the number of stars between the model and observed CMD. %in the regions of the diagram.

We tried fitting several combinations of the four sets of isochrones and found the best-fit solution for a metallicity  \Z=-1.3. The corresponding SFH and model CMD are displayed in the {\em upper} and {\em lower-left} panel of Fig. \ref{sfh}, respectively. The error bars reflect the 1$\sigma$ confidence interval on the star formation rate averaged over an age bin.
According to the recovered history of star formation, the stellar population in NW7 is metal-poor and older than 4 Gyr ago, and  most of the stars formed around 6$^{+1.6}_{-1.3}$ Gyr ago. This agrees with our qualitative analysis in Sect. 5. The derived star formation rate corresponds to a total stellar mass of $\sim 2 \times 10^6$ \msun.
A comparison to the observed data ({\em lower-right} panel) shows a reasonably good match. An isochrone at 6.3 Gyr (log(age) = 9.8) is overlaid on both diagrams to guide the reader's eye. However, the match is not perfect since the colour of the simulated RGB appears roughly 0.1 mag redder than the observed one at $I_0 > 23.5$, %suggesting that the code might be overestimating the number of stars in the oldest bin,
and $\sim 0.05$ mag bluer in the upper part of the RGB. Adding higher metallicity isochrones at \Z = -1 to the model diagram in order to match the upper part of the RGB did not help us to improve the fit but led to an overall too red CMD.

\section{Stellar spatial distribution and comparison to the \hi features}

\begin{figure*}
   \centering
 \includegraphics[width=14.5cm, bb= 0 0 565 450]{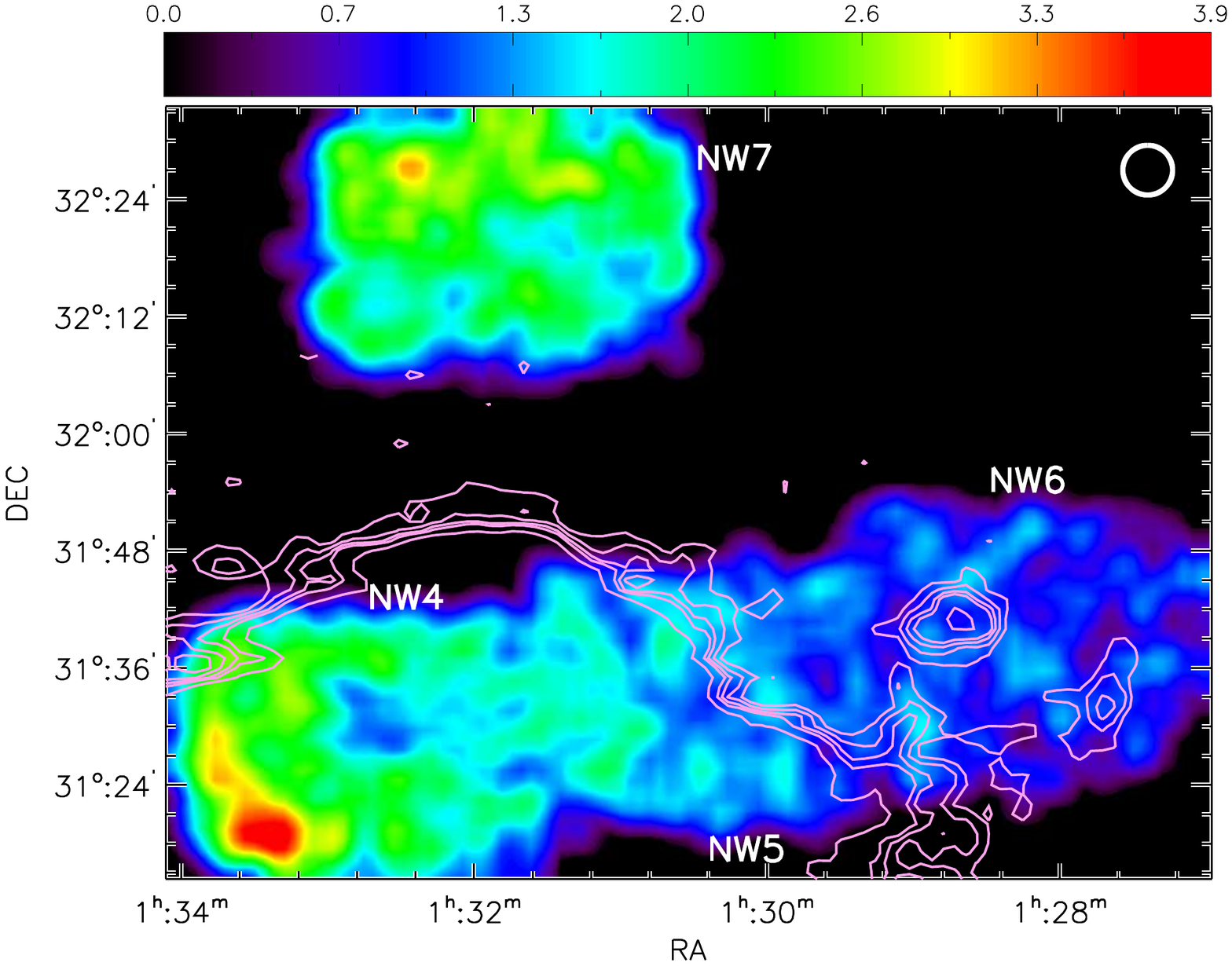}   %fig per astro-ph
   \caption{Spatial distribution of RGB and RC stars with 20.7 mag $< I <$ 24.6 mag in the northern fields. The resolution of the map is 1\armin $\times$ 1\armin, smoothed with a Gaussian kernel of 5 pixels FWHM (shown at the top-right corner of the figure). The colour scale indicates the number of stars per pixel. The density contours of the \hi clouds detected in this region are overlaid as a comparison. Contours are at 3.5, 6, 8.5, 11, 13.5 $\times 10^{18}$ cm$^{-2}$.
   }  \label{dens_map_north}%
    \end{figure*}

Figures \ref{dens_map_south} and \ref{dens_map_north} show the spatial distribution of the RGB and RC stars in the southern and northern fields, respectively.
We selected stars with $I$ magnitudes in the range 20.7 mag $< I <$ 24.6 mag, divided the fields into square bins with a size of 1\armin $\times$ 1\armin, and determined the number of stars in each bin.
The final image of the spatial distribution was obtained after convolving the 2-D histogram with a Gaussian filter of FWHM five times the bin size (5\armin). %No corrections for possible contaminants has been applied (so far!).
The density contours of the main \hi features detected in these regions are overlaid for comparison.

The figures show several structures overlaid on a more diffuse distribution of stars.
The main features in Fig. \ref{dens_map_south} are:
\begin{itemize}

\item The main overdensity at the top-right corner of the figure corresponds to the edge of the optical disc.
A filament-like overdensity is connected to the stellar disc and  extends out to the upper edge of the \hi complex (field SE3).

\item A fainter concentration of stars in the central part of the figure (corresponding to field SE2) at a declination of 30$^{\circ}$, possibly related to the peak density of the main \hi complex. %(field 2)

\item An arc-like feature is visible to the southwest showing an enhancement in the distribution of stars in this region. This overdensity is not, however, spatially coincident with the southern extension of the \hi complex.  It is possible that this increase in the number density of stars might be related instead to the stellar counterpart of the southern \hi warp \citep{2009Natur.461...66M}.

\end{itemize}

 In the northern fields, an extended stellar structure is detected in NW4 and NW5, along the \hi warp. Most of the stars appear to be located within the \hi disc, beyond which the stellar density rapidly decreases. Isolated and fainter clumps can be identified in field NW6. In particular, two overdensities match the position of one of the northern \hi clouds (whose column density contours are overlaid in the figure), although the stellar overdensities appear to be offset from the peak of the \hi column density.
Figure \ref{dens_map_north} mainly stresses the difference in the stellar distribution between the fields, along the \hi warp and the northernmost region. While the stellar density decreases from NW4 to NW6, field NW7 shows a higher concentration of stars, especially to the northern side of the field where few overdensities are visible. The distribution of stars stretches to the north-west towards M31 forming a low surface brightness structure  that extends to a maximum projected radius of $\sim$ 3 degrees as one can see from the map of \citet{2010ApJ...723.1038M}.

%________________________________________________________________

\section{Discussion}

\subsection{\hi clouds and stellar counterparts}

\begin{figure*}
   \centering
 \includegraphics[width=9cm,bb= 5 5 600 430]{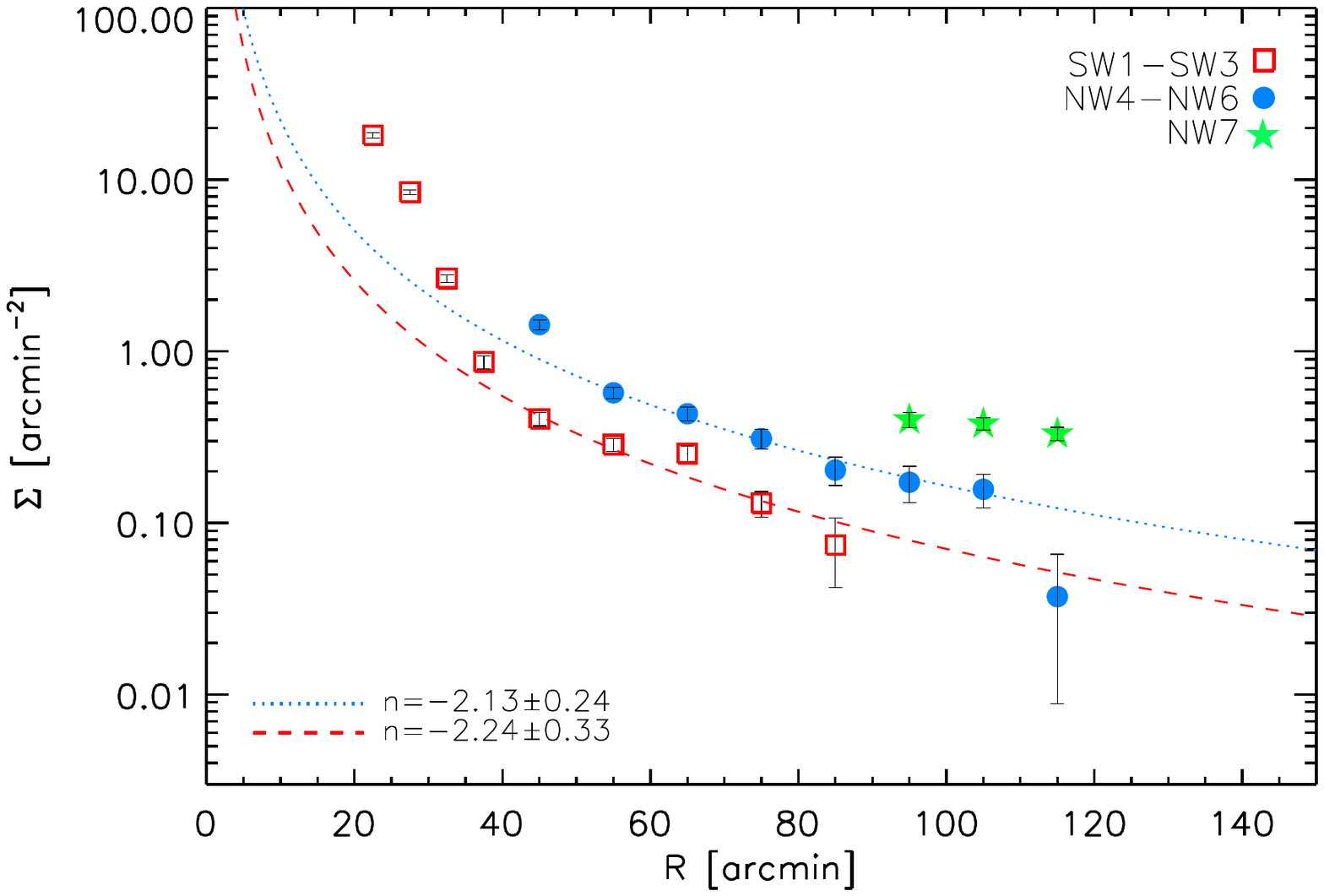} %fig grande
 \includegraphics[width=9cm,bb= 5 5 600 430]{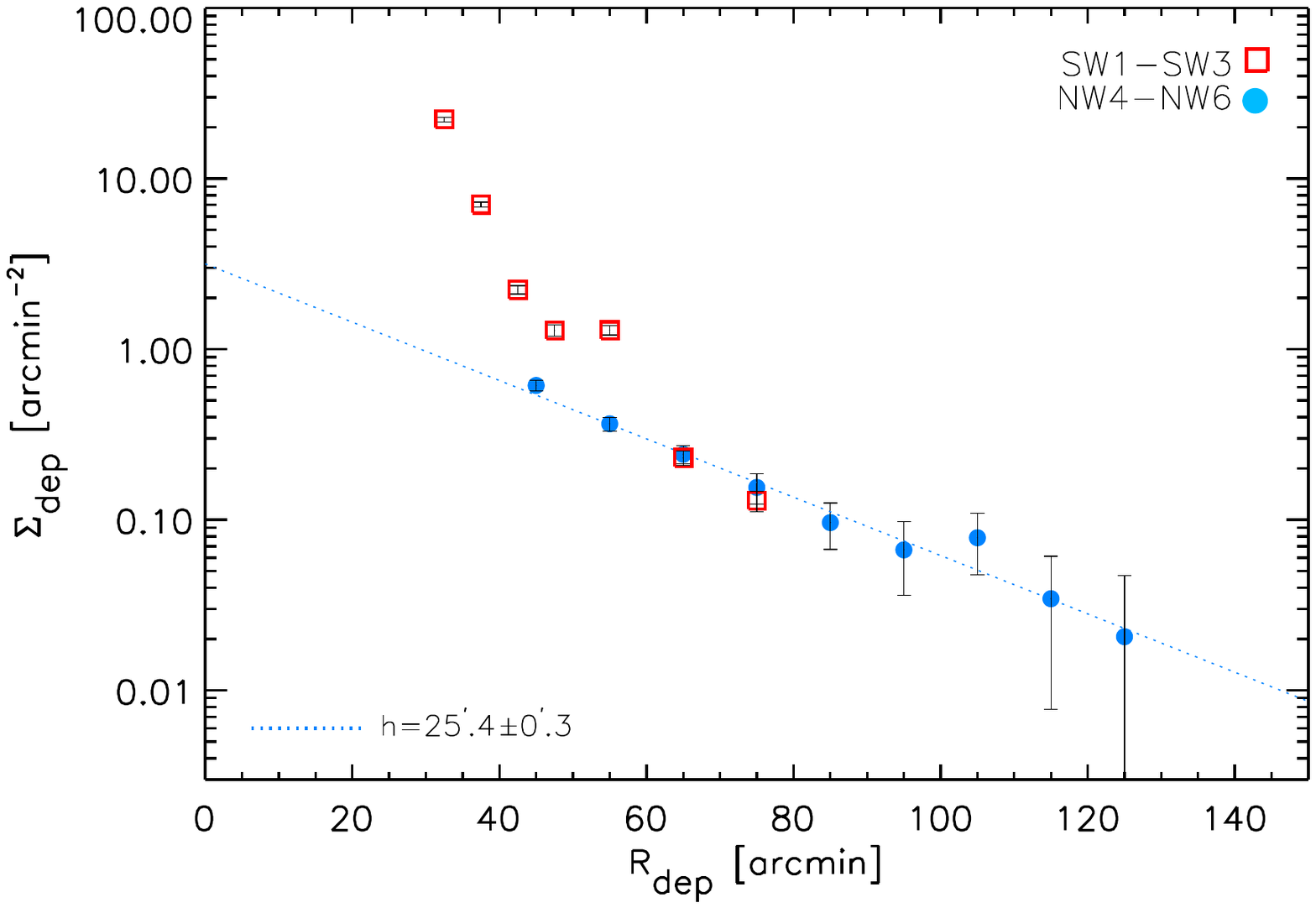}   %fig per astro-ph
   \caption{Stellar surface density profiles in the observed fields versus projected ({\em left}) and deprojected ({\em right}) radius. The open squares and filled dots show the stellar surface density in the SE1-SE3, and NW4-NW6 fields, respectively. Star symbols indicate the density at large radii when field NW7 is also included. The M33 stellar disc extends out to 40$^{\prime}$, while at larger radii there is a clear change in the surface density.
   A power-law profile $(R^{n})$ is fitted to the data in the {\em left} panel to investigate the presence of a stellar halo. We did not attempt to fit the star points because of the anomalous overdensity in field NW7.
   An extended disc component (e$^{-R/h}$) is fitted to the data in the {\em right} panel.
   The best-fit parameters for each model are displayed in the panels.}
   \label{stellar_dens}%
    \end{figure*}

The search for optical counterparts to \hi clouds has often turned out to be
a difficult task with very few successful detections. An optical study of 264 high
velocity clouds down to a surface brightness limit of $\mu_V \sim 26$ mag arcsec$^{-2}$ resulted in no detections
\citep{2002ApJ...574..726S}.
Even for one of the most massive high velocity clouds in the Milky Way,
complex H \citep{1971A&A....14..489H,1971A&A....12...59D}, infrared, and millimeter wavelength observations did not reveal any evidence of either current star formation or an evolved population \citep{2006ApJ...640..270S}.
However, \hi clouds are often found close to stellar streams or disturbances in the stellar disc, as seems to be the case in M33. For example, the \hi clouds found near the southeastern edge of the disc of M31
\citep{2004ApJ...601L..39T,2005A&A...436..101W,2008MNRAS.390.1691W} are spatially and kinematically
correlated to the giant stellar stream \citep{2001Natur.412...49I}, implying that those clouds could be the result of either tidal or ram-pressure stripping of a former satellite galaxy.

The main purpose of our study has been to search for an optical counterpart to the most extended \hi features that
we found in a 21-cm survey around M33.
The cloud properties seems to be indicative of a process of accretion or tidal disturbance in the outer regions of M33.
This is confirmed by the features in the stellar population discussed in this paper and previous works, and also by
the discovery of an extended cluster to the southeast of the galaxy \citep{2008AJ....135.1482S}.
However we do not find convincing evidence of a young stellar population ($\sim$ 100 Myr) related to these features.
This means that there has not been in situ star formation in these clouds, as indicated by their low column
densities, whose peak values in \hi are only around 10$^{19}$ cm$^{-2}$ (Grossi et al. 2008).
We find a sparse and more evolved stellar population in the proximity of the \hi complexes in SE1-SE2,
and some isolated clumps of stars are visible throughout the outer field NW6, but there is no clear evidence of a
correlation between the stellar structures and the gas clouds.

On the other hand, the stellar structures detected in fields NW4 and NW5 (Fig. \ref{dens_map_north}) follow the distribution of the \hi warp, suggesting that these stars represent the optical counterpart to the perturbed
\hi disc, and that the stellar component -- as in the case of the gaseous one -- has a warped distribution.

\subsection{Stellar surface density at large radii: stellar halo or extended disc?}

The fields we observed with Subaru/Suprime-Cam probe different regions beyond the optical disc of M33 ($R_{25} =$
35\arcmin.5).
Combining the information gathered from the analysis of the stellar populations, we found a trend of decreasing
metallicity with radius.
Determining the radial profile of the stellar surface density can help us to discriminate the nature of the different
stellar components that are related to these metallicity variations at large radii
(extended disc or a stellar halo).
Thus, to derive the surface density we divided the surveyed area into radial bins using two different
geometries: $a)$ circular rings to investigate the possibility
that most of the stars we detect trace a smooth stellar halo;
$b)$ elliptical annuli with position angle and inclination given by \citet{2000MNRAS.311..441C}, if
most of the stars lie in an
extended disc that follows the \hi distribution. We discuss these possibilities
in the following subsections.

We selected RGB stars with $I_0$ between 20.7 mag  and 24 mag, and colours \vi $>$ 0.7 mag,
limiting our analysis to RGB stars to ensure a higher completeness level on the photometry.

\subsubsection{Stellar halo density profile}

We derived the number counts in radial circular bins of width 5\arminsp for $R <$ 40\arminsp and width 10\arminsp for $R >$ 40\armin. A different spacing was chosen to more accurately trace the profile at the edge of the M33 stellar disc.
The number counts in each bin were then divided by the area of the corresponding region. The level of contamination was estimated from the control field selecting the number of detections within the same magnitude and colour range of the RGB stars (see Fig. \ref{cmdbg}). The surface density of contaminants was found to be $\sim$ 0.1 arcmin$^{-2}$, and subtracted from the number counts in each radial bin. %The final result is shown in fig. \ref{stellar_dens}.

Figure \ref{stellar_dens} ({\em left} panel) shows the RGB surface density versus the projected radius.
The open squares and the filled dots indicate the contamination-subtracted surface density in SE1-SE3 and NW4-NW6, respectively. Star symbols represent the surface density when NW7 is also taken into account. Error bars are given by the square root of involved star counts scaled by the area of each region.

Our data set allows us to study the decline of the M33 density profile out to a projected radius of 90\arminsp and 120$^{\prime}$ in the SE and NW fields, respectively.
A change in the shape of the profile occurs in the SE fields (open squares) at radii greater than 40$^{\prime}$, where the optical disc ends; beyond 50\armin, the profile shows a similar trend in both the northern and southern regions.
The density enhancement at radii greater than 90$^{\prime}$ (star symbols) is due to the population in field NW7 belonging to the large stellar feature to the north of M33. The density here is about two times higher than that found in NW6 only. % because of %showing the overdensity associated to
%the large stellar feature around M33 partially covered by this field.
The overdensity extends further north of NW7 for about one degree along the direction to M31 (McConnachie et al. 2010).

%To investigate the nature of the stellar distribution
A power-law profile, $R^{n}$, representing the spatial distribution of a spheroidal stellar halo, was then fitted to the data at $R >$ 40\arminsp, leaving out the overdensity in NW7 (star symbols).
We find best-fit power-law indices $n = -2.13 \pm 0.24$ and $n = -2.24 \pm 0.33$ in the NW and SE regions, with a $\chi^2$ of 1.4 and 1.8 , respectively.
The two resulting power-law indices are compatible with a unique distribution,
even though the northern side has a somewhat higher density.
Our result is comparable to that found in the MW and M31 stellar halos:
the surface density profile of halo stars in the Galaxy falls off with galactic radius following a power-law index varying between -1.5 and -2.6 \citep{2000AJ....119.2843C,2008ApJ...673..864J,2008ApJ...680..295B,2008ApJ...673..864J}. %The most recent SDSS study by \citet{} found $n = -1.8$.
In M31, Ibata et al. (2007)  %in their analysis of the M31 along the southern extent of its minor axis
%fitted the radial surface density with a Hernquist profile with $r_s \sim 53$ kpc or an exponential profile with scale-length $h \sim 47$ kpc out to 150 kpc. A
obtained a surface density profile with a power-law index of $n = -1.91 \pm 0.12$, while \citet{2010ApJ...708.1168T}  found
a steeper power law with $n$ = -2.17 $\pm$ 0.15.
%, given their different properties

\subsubsection{An extended outer disc}

We investigate the presence of an extended stellar disc under the assumption that it follows the same orientation as
the warped \hi disc.
If an extended stellar disc is in place, the stellar surface density %perpendicular
%to the galactic plane
can be inferred by deprojecting the stars onto the plane of the
galaxy. For a warped disc, a tilted ring model is needed. Following \citet{2000MNRAS.311..441C}, we assumed that the
outer disc
position angle varies between 21$^{\circ}$ and -11$^{\circ}$, while its inclination
stays roughly constant at about 50$^{\circ}$. Hence, we computed the location of
the elliptical contours in the plane of the sky corresponding to fixed
galactocentric radii, i.e. the radii of the tilted rings. We used radial spacing of 5\arminsp for the innermost ellipses
and 10\arminsp for ellipses with major axis larger than 50\armin. We then assigned each star to a ring and compute
the stellar
surface density in the ring areas covered by our survey.

The deprojected surface density is displayed in the {\em right} panel of Fig. \ref{stellar_dens}.
The figure now shows an excess density in the SE fields between 40\arminsp and 60\arminsp,
and a very good match between the density values of the northern and southern regions in the following two bins
(60\arminsp $< R_{dep} <$ 80\armin).
An exponential profile was fitted to the density  distribution for $R_{dep} >$ 40\arminsp. The best-fit scale-length  is  $h = $25\armin.4 $\pm$ 0\armin.3, with $\chi^2 =$ 0.7.
This scale-length ($\sim$ 7 kpc) is much larger than the stellar scale-length
of the inner disc ($\sim$ 1.5 kpc). The stellar drop-off in M33 is also slightly shallower than the \hi decline in the
same area.

Given the better fit obtained after deprojecting the stellar distribution, we conclude that the observed surface
density in our fields seems to be more consistent with an extended disc component than a smooth stellar halo.
However, we emphasize that our analysis is based on fields selected along the \hi warp, and
our results do not exclude that the contribution of a  faint spheroidal halo
could be investigated more effectively in different regions around M33.
Fields from the PAndAS survey at similar angular distance as NW6 and NW7 along directions not polluted by either the
disc or the large stellar feature around the galaxy, % \citep{2010ApJ...723.1038M},
 contain a small RGB population and are presumed to represent %by McConnachie et al. (2010)
an underlying halo population in this galaxy.

\subsection{The large stellar feature around M33}

The origin of the large stellar feature around M33 is discussed in \citet{2009Natur.461...66M,2010ApJ...723.1038M}, and we refer the reader to those papers for more details.
Here we summarise the main conclusions of these authors in light of our additional results.
Their favoured scenario assumes that the structure is the result of  a tidal disturbance caused by the motion of M33 around M31.  The northwest - southeast symmetry of the "S-shaped" structure, and the rough alignment with the \hi warp, are the main pieces of evidence supporting this interpretation. Simulations of the M31-M33 system %\citep{2009Natur.461...66M}
have showed that a close encounter at a pericentre distance of about 40 kpc would excite tidal tails in M33 without severely distorting or disrupting
the disc. These simulations also set the epoch of the encounter to be between 2 and 3 Gyr ago.
However, %as it has also been pointed out in \citet{2010ApJ...723.1038M},
within this scenario it would be difficult to reconcile the metallicity difference between the structure (\Z=-1.3) and the disc (\Z=-0.7,-0.4) stellar populations, a difference that we also detect within our data set.

On the other hand, if the structure originated from the disruption of a dwarf galaxy such a difference in the metal abundance would not be an issue, and this scenario is also considered as a plausible alternative. The lack of a visible progenitor would imply that the dwarf galaxy has been entirely destroyed prior to the current epoch.  From the SFH that we derived in Sect. 6, the total stellar mass in field NW7 would amount to $\sim 2 \times 10^6$ \msun, and \citet{2010ApJ...723.1038M} estimate that the total luminosity of the whole structure is M$_V = -12.7$, comparable to that of a bright LG dSph.
Our data cannot provide constraints that enable us to rule out one of these possibilities,
but it is interesting to note that the lower limit we infer to the age of the stars (log(age) = 9.5 or $\sim 3.2$ Gyr), is comparable to the epoch of the encounter between the two galaxies predicted by the simulations.

 %________________________________________________________________

\begin{figure}
   \centering
 \includegraphics[width=9cm,bb= 5 5 590 410]{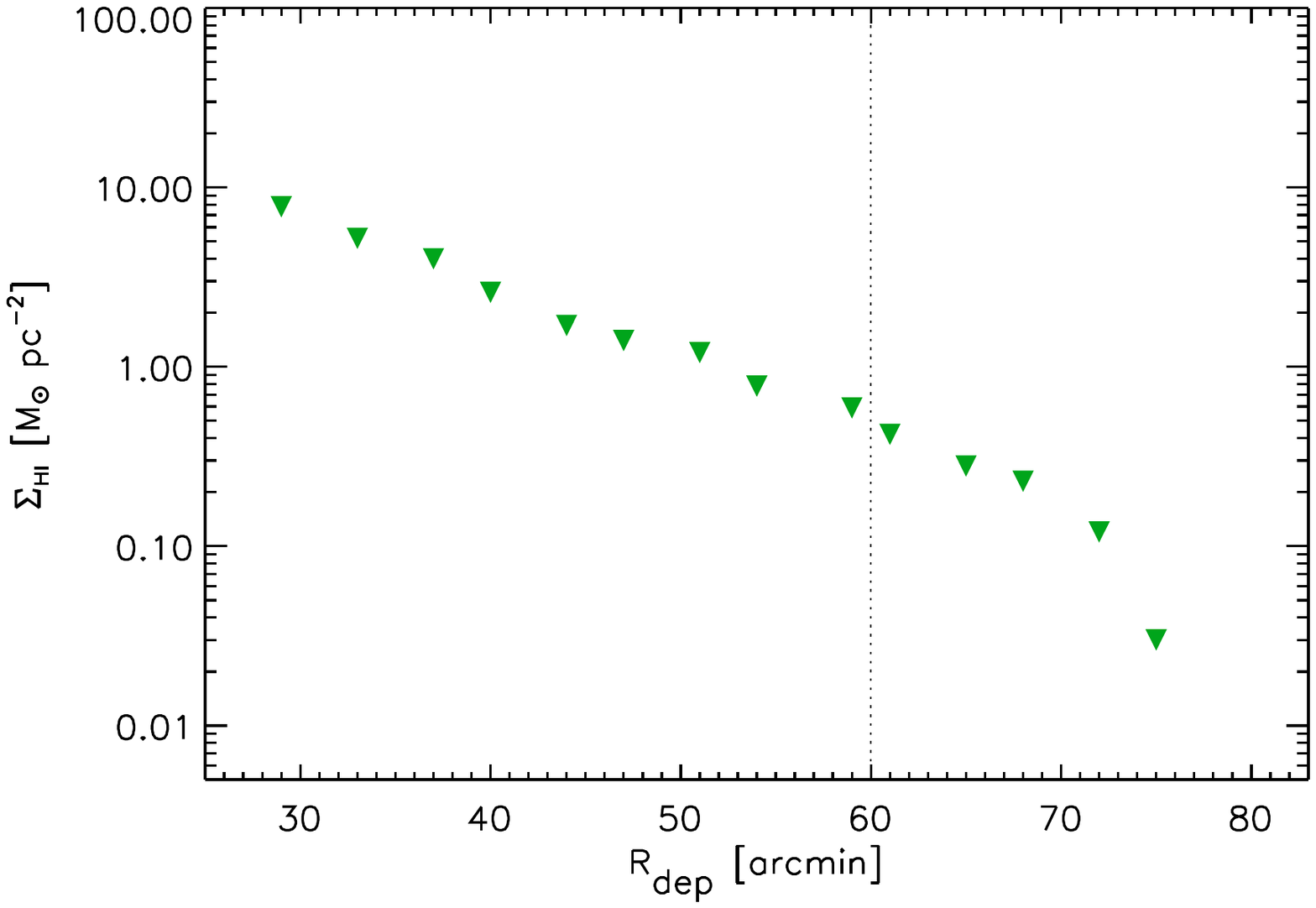} %fig grande
   \caption{\hi surface density versus deprojected radius. The vertical line shows the maximum distance at which young blue stars were detected. Young stars are found out to a radius that is comparable to the distance beyond which the \hi surface density declines abruptly. }
   \label{young}%
    \end{figure}

\subsection{Recent star formation beyond \rsb}

In Sect. 3, we pointed out that a population of young stars, with ages around or older than 200 Myr, was detected in SE3, NW4, and NW5. The majority of them were found within 50\armin, with a few blue stars detected out to a galactocentric radii of 60\armin ($\sim$ 17 kpc; see Fig. \ref{rad_cmd_south} and \ref{rad_cmd_north}). Finding evidence of  recent star formation activity is surprising at these radii, given the low column density of the gas.
As a comparison, we show in Fig. \ref{young} the M33 \hi surface density as a function of the galactocentric distance along the \hi warp.
At 60\armin, the gas surface density is around 0.6 \msun pc$^{-2}$ (or $\sim 7.5 \times 10^{19}$ cm$^{-2}$), roughly eight times lower than the density at \rsb, then starts to decline abruptly at $R_{dep} >$ 70\armin.

The presence of such young populations at large galactocentric radii challenges the current picture of star formation given by the Schmidt-Kennicut law \citep{1998ARA&A..36..189K}. However,
studies of nearby spirals based mainly on GALEX observations have shown that star formation can also occur beyond $R_{25}$ in a low-density environment \citep{2005ApJ...619L..79T,2007ApJS..173..538T}.
Stellar associations with an average age of $\sim$ 150 Myr were found in the outer discs of a sample of five nearby spiral galaxies (Alberts et al. 2011).
\citet{2010AJ....140.1194B} studied star formation rates in the far outer discs of spirals where \hi is the dominant gaseous component. In contrast to the inner parts of the disc, the \hi density and the star formation rate (derived by the far UV emission) showed an overall good correlation, implying that different processes may govern the formation of stars at large galactocentric distances.

\section{Summary}

We have presented Subaru/Suprime-Cam $V$ and $I$ observations of seven fields around the disc of M33. The fields cover regions of the sky where \hi complexes were detected by the ALFALFA survey, the northern \hi warp, and part of the extended feature that was discovered around the galaxy by the PAndAS survey.
We have studied the properties of the different stellar populations using CMDs, estimated the mean metallicity and age in each region, and determined the radial distribution of stars.
Here we summarise our main results:

\begin{enumerate}

\item We confirmed the presence of an extended population of evolved stars (in the RGB and RC phase) with ages between 1 and 10 Gyr  out to an angular distance of 120\arminsp ($\sim$ 30 kpc).

\item A sparse population of young stars with ages around 100-200 Myr is detected both to the NW and to the SE out to a galactocentric radius of $\sim$15 kpc.

\item We did not find a clear optical counterpart to the main \hi complexes detected around the optical disc, although stellar structures are detected in all target regions. This provides evidence that the stellar distribution at large radii is disturbed in a similar way to the gaseous one because of either tidal interaction or accretion.

\item We used the shape and colour of the RGB as well as the magnitude and colour of the RC to derive the metallicity of the dominant stellar population in each field. The mean metallicity gradually decreases from \Z=-0.4 at the edge of the optical disc, down to $\lesssim -1$, out to a radius of $\sim$ 30 kpc.

\item %An extended disc component is detected in both SE and NW regions out to a projected distance of 120\arminsp ($\sim$ 30 kpc).
    The surface radial profile of the stellar distribution in our fields %along the northern fields (NW4, NW5, and NW6)
    can be more closely fitted by an extended disc component with a scale-length of $\sim$ 7 kpc, rather than a
    smooth stellar halo. This suggests that the stars as in the case of the gas have a warped distribution.
    However, this does not exclude the presence of an underlying stellar halo, which may be more accurately studied by targeting other regions around the galaxy.

\item Finally we analysed the stellar populations in a region between 1.5 and 2 degrees north of M33 probing part of the large stellar plume surrounding the galaxy. The RGB and RC colours show a more metal-poor and older population than in the other fields. A CMD fitting analysis shows that stars in this region formed between 3 and 10 Gyr, with a metallicity of \Z = -1.3.

\end{enumerate}

\begin{acknowledgements}
We thank the staff at Subaru telescope for their support during the observing run, and the anonymous referee for the comments and suggestions that contributed to improve the paper.
\end{acknowledgements}

\bibliographystyle{aa} % style aa.bst
\bibliography{M33_streams_bib} % your references Yourfile.bib

\begin{thebibliography}{70}
\expandafter\ifx\csname natexlab\endcsname\relax\def\natexlab#1{#1}\fi

\bibitem[{{Abadi} {et~al.}(2003){Abadi}, {Navarro}, {Steinmetz}, \&
  {Eke}}]{2003ApJ...591..499A}
{Abadi}, M.~G., {Navarro}, J.~F., {Steinmetz}, M., \& {Eke}, V.~R. 2003, \apj,
  591, 499

\bibitem[{{Bekki}(2008)}]{2008MNRAS.390L..24B}
{Bekki}, K. 2008, \mnras, 390, L24

\bibitem[{{Bell} {et~al.}(2008){Bell}, {Zucker}, {Belokurov}, {Sharma},
  {Johnston}, {Bullock}, {Hogg}, {Jahnke}, {de Jong}, {Beers}, {Evans},
  {Grebel}, {Ivezi{\'c}}, {Koposov}, {Rix}, {Schneider}, {Steinmetz}, \&
  {Zolotov}}]{2008ApJ...680..295B}
{Bell}, E.~F., {Zucker}, D.~B., {Belokurov}, V., {et~al.} 2008, \apj, 680, 295

\bibitem[{{Bellazzini} {et~al.}(2001){Bellazzini}, {Ferraro}, \&
  {Pancino}}]{2001ApJ...556..635B}
{Bellazzini}, M., {Ferraro}, F.~R., \& {Pancino}, E. 2001, \apj, 556, 635

\bibitem[{{Belokurov} {et~al.}(2007){Belokurov}, {Evans}, {Irwin},
  {Lynden-Bell}, {Yanny}, {Vidrih}, {Gilmore}, {Seabroke}, {Zucker},
  {Wilkinson}, {Hewett}, {Bramich}, {Fellhauer}, {Newberg}, {Wyse}, {Beers},
  {Bell}, {Barentine}, {Brinkmann}, {Cole}, {Pan}, \&
  {York}}]{2007ApJ...658..337B}
{Belokurov}, V., {Evans}, N.~W., {Irwin}, M.~J., {et~al.} 2007, \apj, 658, 337

\bibitem[{{Belokurov} {et~al.}(2006){Belokurov}, {Zucker}, {Evans}, {Gilmore},
  {Vidrih}, {Bramich}, {Newberg}, {Wyse}, {Irwin}, {Fellhauer}, {Hewett},
  {Walton}, {Wilkinson}, {Cole}, {Yanny}, {Rockosi}, {Beers}, {Bell},
  {Brinkmann}, {Ivezi{\'c}}, \& {Lupton}}]{2006ApJ...642L.137B}
{Belokurov}, V., {Zucker}, D.~B., {Evans}, N.~W., {et~al.} 2006, \apjl, 642,
  L137

\bibitem[{{Bigiel} {et~al.}(2010){Bigiel}, {Leroy}, {Walter}, {Blitz},
  {Brinks}, {de Blok}, \& {Madore}}]{2010AJ....140.1194B}
{Bigiel}, F., {Leroy}, A., {Walter}, F., {et~al.} 2010, \aj, 140, 1194

\bibitem[{{Brook} {et~al.}(2010){Brook}, {Governato}, {Roskar}, {Stinson},
  {Brooks}, {Wadsley}, {Quinn}, {Gibson}, {Snaith}, {Pilkington}, \&
  {House}}]{2010arXiv1010.1004B}
{Brook}, C.~B., {Governato}, F., {Roskar}, R., {et~al.} 2010, ArXiv e-prints

\bibitem[{{Brooks} {et~al.}(2009){Brooks}, {Governato}, {Quinn}, {Brook}, \&
  {Wadsley}}]{2009ApJ...694..396B}
{Brooks}, A.~M., {Governato}, F., {Quinn}, T., {Brook}, C.~B., \& {Wadsley}, J.
  2009, \apj, 694, 396

\bibitem[{{Brooks} {et~al.}(2004){Brooks}, {Wilson}, \&
  {Harris}}]{2004AJ....128..237B}
{Brooks}, R.~S., {Wilson}, C.~D., \& {Harris}, W.~E. 2004, \aj, 128, 237

\bibitem[{{Bullock} \& {Johnston}(2005)}]{2005ApJ...635..931B}
{Bullock}, J.~S. \& {Johnston}, K.~V. 2005, \apj, 635, 931

\bibitem[{{Chiba} \& {Beers}(2000)}]{2000AJ....119.2843C}
{Chiba}, M. \& {Beers}, T.~C. 2000, \aj, 119, 2843

\bibitem[{{Ciardullo} {et~al.}(2004){Ciardullo}, {Durrell}, {Laychak},
  {Herrmann}, {Moody}, {Jacoby}, \& {Feldmeier}}]{2004ApJ...614..167C}
{Ciardullo}, R., {Durrell}, P.~R., {Laychak}, M.~B., {et~al.} 2004, \apj, 614,
  167

\bibitem[{{Corbelli} \& {Salucci}(2000)}]{2000MNRAS.311..441C}
{Corbelli}, E. \& {Salucci}, P. 2000, \mnras, 311, 441

\bibitem[{{Corbelli} \& {Schneider}(1997)}]{1997ApJ...479..244C}
{Corbelli}, E. \& {Schneider}, S.~E. 1997, \apj, 479, 244

\bibitem[{{Dekel} \& {Birnboim}(2006)}]{2006MNRAS.368....2D}
{Dekel}, A. \& {Birnboim}, Y. 2006, \mnras, 368, 2

\bibitem[{{Dieter}(1971)}]{1971A&A....12...59D}
{Dieter}, N.~H. 1971, \aap, 12, 59

\bibitem[{{Durrell} {et~al.}(2010){Durrell}, {Sarajedini}, \&
  {Chandar}}]{2010ApJ...718.1118D}
{Durrell}, P.~R., {Sarajedini}, A., \& {Chandar}, R. 2010, \apj, 718, 1118

\bibitem[{{Fardal} {et~al.}(2008){Fardal}, {Babul}, {Guhathakurta}, {Gilbert},
  \& {Dodge}}]{2008ApJ...682L..33F}
{Fardal}, M.~A., {Babul}, A., {Guhathakurta}, P., {Gilbert}, K.~M., \& {Dodge},
  C. 2008, \apjl, 682, L33

\bibitem[{{Ferguson} {et~al.}(2002){Ferguson}, {Irwin}, {Ibata}, {Lewis}, \&
  {Tanvir}}]{2002AJ....124.1452F}
{Ferguson}, A.~M.~N., {Irwin}, M.~J., {Ibata}, R.~A., {Lewis}, G.~F., \&
  {Tanvir}, N.~R. 2002, \aj, 124, 1452

\bibitem[{{Font} {et~al.}(2006){Font}, {Johnston}, {Guhathakurta}, {Majewski},
  \& {Rich}}]{2006AJ....131.1436F}
{Font}, A.~S., {Johnston}, K.~V., {Guhathakurta}, P., {Majewski}, S.~R., \&
  {Rich}, R.~M. 2006, \aj, 131, 1436

\bibitem[{{Freedman} {et~al.}(2001){Freedman}, {Madore}, {Gibson}, {Ferrarese},
  {Kelson}, {Sakai}, {Mould}, {Kennicutt}, {Ford}, {Graham}, {Huchra},
  {Hughes}, {Illingworth}, {Macri}, \& {Stetson}}]{2001ApJ...553...47F}
{Freedman}, W.~L., {Madore}, B.~F., {Gibson}, B.~K., {et~al.} 2001, \apj, 553,
  47

\bibitem[{{Giovanelli} {et~al.}(2005){Giovanelli}, {Haynes}, {Kent},
  {Perillat}, {Saintonge}, {Brosch}, {Catinella}, {Hoffman}, {Stierwalt},
  {Spekkens}, {Lerner}, {Masters}, {Momjian}, {Rosenberg}, {Springob},
  {Boselli}, {Charmandaris}, {Darling}, {Davies}, {Garcia Lambas}, {Gavazzi},
  {Giovanardi}, {Hardy}, {Hunt}, {Iovino}, {Karachentsev}, {Karachentseva},
  {Koopmann}, {Marinoni}, {Minchin}, {Muller}, {Putman}, {Pantoja}, {Salzer},
  {Scodeggio}, {Skillman}, {Solanes}, {Valotto}, {van Driel}, \& {van
  Zee}}]{2005AJ....130.2598G}
{Giovanelli}, R., {Haynes}, M.~P., {Kent}, B.~R., {et~al.} 2005, \aj, 130, 2598

\bibitem[{{Girardi} \& {Salaris}(2001)}]{2001MNRAS.323..109G}
{Girardi}, L. \& {Salaris}, M. 2001, \mnras, 323, 109

\bibitem[{{Governato} {et~al.}(2007){Governato}, {Willman}, {Mayer}, {Brooks},
  {Stinson}, {Valenzuela}, {Wadsley}, \& {Quinn}}]{2007MNRAS.374.1479G}
{Governato}, F., {Willman}, B., {Mayer}, L., {et~al.} 2007, \mnras, 374, 1479

\bibitem[{{Grillmair}(2006)}]{2006ApJ...645L..37G}
{Grillmair}, C.~J. 2006, \apjl, 645, L37

\bibitem[{{Grillmair}(2009)}]{2009ApJ...693.1118G}
{Grillmair}, C.~J. 2009, \apj, 693, 1118

\bibitem[{{Grossi} {et~al.}(2008){Grossi}, {Giovanardi}, {Corbelli},
  {Giovanelli}, {Haynes}, {Martin}, {Saintonge}, \&
  {Dowell}}]{2008A&A...487..161G}
{Grossi}, M., {Giovanardi}, C., {Corbelli}, E., {et~al.} 2008, \aap, 487, 161

\bibitem[{{Harris} \& {Zaritsky}(2001)}]{2001ApJS..136...25H}
{Harris}, J. \& {Zaritsky}, D. 2001, \apjs, 136, 25

\bibitem[{{Hulsbosch}(1971)}]{1971A&A....14..489H}
{Hulsbosch}, A.~N.~M. 1971, \aap, 14, 489

\bibitem[{{Ibata} {et~al.}(2001){Ibata}, {Irwin}, {Lewis}, {Ferguson}, \&
  {Tanvir}}]{2001Natur.412...49I}
{Ibata}, R., {Irwin}, M., {Lewis}, G., {Ferguson}, A.~M.~N., \& {Tanvir}, N.
  2001, \nat, 412, 49

\bibitem[{{Ibata} {et~al.}(1994){Ibata}, {Gilmore}, \&
  {Irwin}}]{1994Natur.370..194I}
{Ibata}, R.~A., {Gilmore}, G., \& {Irwin}, M.~J. 1994, \nat, 370, 194

\bibitem[{{Juri{\'c}} {et~al.}(2008){Juri{\'c}}, {Ivezi{\'c}}, {Brooks},
  {Lupton}, {Schlegel}, {Finkbeiner}, {Padmanabhan}, {Bond}, {Sesar},
  {Rockosi}, {Knapp}, {Gunn}, {Sumi}, {Schneider}, {Barentine}, {Brewington},
  {Brinkmann}, {Fukugita}, {Harvanek}, {Kleinman}, {Krzesinski}, {Long},
  {Neilsen}, {Nitta}, {Snedden}, \& {York}}]{2008ApJ...673..864J}
{Juri{\'c}}, M., {Ivezi{\'c}}, {\v Z}., {Brooks}, A., {et~al.} 2008, \apj, 673,
  864

\bibitem[{{Kennicutt}(1998)}]{1998ARA&A..36..189K}
{Kennicutt}, Jr., R.~C. 1998, \araa, 36, 189

\bibitem[{{Kim} {et~al.}(2002){Kim}, {Kim}, {Lee}, {Sarajedini}, \&
  {Geisler}}]{2002AJ....123..244K}
{Kim}, M., {Kim}, E., {Lee}, M.~G., {Sarajedini}, A., \& {Geisler}, D. 2002,
  \aj, 123, 244

\bibitem[{{Landolt}(1992)}]{1992AJ....104..340L}
{Landolt}, A.~U. 1992, \aj, 104, 340

\bibitem[{{Lee} {et~al.}(1993){Lee}, {Freedman}, \&
  {Madore}}]{1993ApJ...417..553L}
{Lee}, M.~G., {Freedman}, W.~L., \& {Madore}, B.~F. 1993, \apj, 417, 553

\bibitem[{{Lee} {et~al.}(2002){Lee}, {Kim}, {Sarajedini}, {Geisler}, \&
  {Gieren}}]{2002ApJ...565..959L}
{Lee}, M.~G., {Kim}, M., {Sarajedini}, A., {Geisler}, D., \& {Gieren}, W. 2002,
  \apj, 565, 959

\bibitem[{{Loeb} {et~al.}(2005){Loeb}, {Reid}, {Brunthaler}, \&
  {Falcke}}]{2005ApJ...633..894L}
{Loeb}, A., {Reid}, M.~J., {Brunthaler}, A., \& {Falcke}, H. 2005, \apj, 633,
  894

\bibitem[{{Madore} \& {Freedman}(1995)}]{1995AJ....109.1645M}
{Madore}, B.~F. \& {Freedman}, W.~L. 1995, \aj, 109, 1645

\bibitem[{{Marigo} {et~al.}(2008){Marigo}, {Girardi}, {Bressan}, {Groenewegen},
  {Silva}, \& {Granato}}]{2008A&A...482..883M}
{Marigo}, P., {Girardi}, L., {Bressan}, A., {et~al.} 2008, \aap, 482, 883

\bibitem[{{McConnachie} {et~al.}(2006){McConnachie}, {Chapman}, {Ibata},
  {Ferguson}, {Irwin}, {Lewis}, {Tanvir}, \& {Martin}}]{2006ApJ...647L..25M}
{McConnachie}, A.~W., {Chapman}, S.~C., {Ibata}, R.~A., {et~al.} 2006, \apjl,
  647, L25

\bibitem[{{McConnachie} {et~al.}(2010){McConnachie}, {Ferguson}, {Irwin},
  {Dubinski}, {Widrow}, {Dotter}, {Ibata}, \& {Lewis}}]{2010ApJ...723.1038M}
{McConnachie}, A.~W., {Ferguson}, A.~M.~N., {Irwin}, M.~J., {et~al.} 2010,
  \apj, 723, 1038

\bibitem[{{McConnachie} {et~al.}(2004){McConnachie}, {Irwin}, {Ferguson},
  {Ibata}, {Lewis}, \& {Tanvir}}]{2004MNRAS.350..243M}
{McConnachie}, A.~W., {Irwin}, M.~J., {Ferguson}, A.~M.~N., {et~al.} 2004,
  \mnras, 350, 243

\bibitem[{{McConnachie} {et~al.}(2009){McConnachie}, {Irwin}, {Ibata},
  {Dubinski}, {Widrow}, {Martin}, {C{\^o}t{\'e}}, {Dotter}, {Navarro},
  {Ferguson}, {Puzia}, {Lewis}, {Babul}, {Barmby}, {Bienaym{\'e}}, {Chapman},
  {Cockcroft}, {Collins}, {Fardal}, {Harris}, {Huxor}, {Mackey},
  {Pe{\~n}arrubia}, {Rich}, {Richer}, {Siebert}, {Tanvir}, {Valls-Gabaud}, \&
  {Venn}}]{2009Natur.461...66M}
{McConnachie}, A.~W., {Irwin}, M.~J., {Ibata}, R.~A., {et~al.} 2009, \nat, 461,
  66

\bibitem[{{Miyazaki} {et~al.}(2002){Miyazaki}, {Komiyama}, {Sekiguchi},
  {Okamura}, {Doi}, {Furusawa}, {Hamabe}, {Imi}, {Kimura}, {Nakata}, {Okada},
  {Ouchi}, {Shimasaku}, {Yagi}, \& {Yasuda}}]{2002PASJ...54..833M}
{Miyazaki}, S., {Komiyama}, Y., {Sekiguchi}, M., {et~al.} 2002, \pasj, 54, 833

\bibitem[{{Newton} \& {Emerson}(1977)}]{1977MNRAS.181..573N}
{Newton}, K. \& {Emerson}, D.~T. 1977, \mnras, 181, 573

\bibitem[{{Ouchi} {et~al.}(2004){Ouchi}, {Shimasaku}, {Okamura}, {Furusawa},
  {Kashikawa}, {Ota}, {Doi}, {Hamabe}, {Kimura}, {Komiyama}, {Miyazaki},
  {Miyazaki}, {Nakata}, {Sekiguchi}, {Yagi}, \& {Yasuda}}]{2004ApJ...611..660O}
{Ouchi}, M., {Shimasaku}, K., {Okamura}, S., {et~al.} 2004, \apj, 611, 660

\bibitem[{{Putman} {et~al.}(2009){Putman}, {Peek}, {Muratov}, {Gnedin}, {Hsu},
  {Douglas}, {Heiles}, {Stanimirovic}, {Korpela}, \&
  {Gibson}}]{2009ApJ...703.1486P}
{Putman}, M.~E., {Peek}, J.~E.~G., {Muratov}, A., {et~al.} 2009, \apj, 703,
  1486

\bibitem[{{Regan} \& {Vogel}(1994)}]{1994ApJ...434..536R}
{Regan}, M.~W. \& {Vogel}, S.~N. 1994, \apj, 434, 536

\bibitem[{{Salpeter}(1955)}]{1955ApJ...121..161S}
{Salpeter}, E.~E. 1955, \apj, 121, 161

\bibitem[{{Sarajedini} {et~al.}(2006){Sarajedini}, {Barker}, {Geisler},
  {Harding}, \& {Schommer}}]{2006AJ....132.1361S}
{Sarajedini}, A., {Barker}, M.~K., {Geisler}, D., {Harding}, P., \& {Schommer},
  R. 2006, \aj, 132, 1361

\bibitem[{{Sarajedini} {et~al.}(2000){Sarajedini}, {Geisler}, {Schommer}, \&
  {Harding}}]{2000AJ....120.2437S}
{Sarajedini}, A., {Geisler}, D., {Schommer}, R., \& {Harding}, P. 2000, \aj,
  120, 2437

\bibitem[{{Schlegel} {et~al.}(1998){Schlegel}, {Finkbeiner}, \&
  {Davis}}]{1998ApJ...500..525S}
{Schlegel}, D.~J., {Finkbeiner}, D.~P., \& {Davis}, M. 1998, \apj, 500, 525

\bibitem[{{Simon} \& {Blitz}(2002)}]{2002ApJ...574..726S}
{Simon}, J.~D. \& {Blitz}, L. 2002, \apj, 574, 726

\bibitem[{{Simon} {et~al.}(2006){Simon}, {Blitz}, {Cole}, {Weinberg}, \&
  {Cohen}}]{2006ApJ...640..270S}
{Simon}, J.~D., {Blitz}, L., {Cole}, A.~A., {Weinberg}, M.~D., \& {Cohen}, M.
  2006, \apj, 640, 270

\bibitem[{{Stetson}(1987)}]{1987PASP...99..191S}
{Stetson}, P.~B. 1987, \pasp, 99, 191

\bibitem[{{Stetson}(1994)}]{1994PASP..106..250S}
{Stetson}, P.~B. 1994, \pasp, 106, 250

\bibitem[{{Stonkut{\'e}} {et~al.}(2008){Stonkut{\'e}}, {Vansevi{\v c}ius},
  {Arimoto}, {Hasegawa}, {Narbutis}, {Tamura}, {Jablonka}, {Ohta}, \&
  {Yamada}}]{2008AJ....135.1482S}
{Stonkut{\'e}}, R., {Vansevi{\v c}ius}, V., {Arimoto}, N., {et~al.} 2008, \aj,
  135, 1482

\bibitem[{{Tanaka} {et~al.}(2010){Tanaka}, {Chiba}, {Komiyama}, {Guhathakurta},
  {Kalirai}, \& {Iye}}]{2010ApJ...708.1168T}
{Tanaka}, M., {Chiba}, M., {Komiyama}, Y., {et~al.} 2010, \apj, 708, 1168

\bibitem[{{Thilker} {et~al.}(2005{\natexlab{a}}){Thilker}, {Bianchi},
  {Boissier}, {Gil de Paz}, {Madore}, {Martin}, {Meurer}, {Neff}, {Rich},
  {Schiminovich}, {Seibert}, {Wyder}, {Barlow}, {Byun}, {Donas}, {Forster},
  {Friedman}, {Heckman}, {Jelinsky}, {Lee}, {Malina}, {Milliard}, {Morrissey},
  {Siegmund}, {Small}, {Szalay}, \& {Welsh}}]{2005ApJ...619L..79T}
{Thilker}, D.~A., {Bianchi}, L., {Boissier}, S., {et~al.} 2005{\natexlab{a}},
  \apjl, 619, L79

\bibitem[{{Thilker} {et~al.}(2007){Thilker}, {Bianchi}, {Meurer}, {Gil de Paz},
  {Boissier}, {Madore}, {Boselli}, {Ferguson}, {Mu{\~n}oz-Mateos}, {Madsen},
  {Hameed}, {Overzier}, {Forster}, {Friedman}, {Martin}, {Morrissey}, {Neff},
  {Schiminovich}, {Seibert}, {Small}, {Wyder}, {Donas}, {Heckman}, {Lee},
  {Milliard}, {Rich}, {Szalay}, {Welsh}, \& {Yi}}]{2007ApJS..173..538T}
{Thilker}, D.~A., {Bianchi}, L., {Meurer}, G., {et~al.} 2007, \apjs, 173, 538

\bibitem[{{Thilker} {et~al.}(2004){Thilker}, {Braun}, {Walterbos}, {Corbelli},
  {Lockman}, {Murphy}, \& {Maddalena}}]{2004ApJ...601L..39T}
{Thilker}, D.~A., {Braun}, R., {Walterbos}, R.~A.~M., {et~al.} 2004, \apjl,
  601, L39

\bibitem[{{Thilker} {et~al.}(2005{\natexlab{b}}){Thilker}, {Hoopes}, {Bianchi},
  {Boissier}, {Rich}, {Seibert}, {Friedman}, {Rey}, {Buat}, {Barlow}, {Byun},
  {Donas}, {Forster}, {Heckman}, {Jelinsky}, {Lee}, {Madore}, {Malina},
  {Martin}, {Milliard}, {Morrissey}, {Neff}, {Schiminovich}, {Siegmund},
  {Small}, {Szalay}, {Welsh}, \& {Wyder}}]{2005ApJ...619L..67T}
{Thilker}, D.~A., {Hoopes}, C.~G., {Bianchi}, L., {et~al.} 2005{\natexlab{b}},
  \apjl, 619, L67

\bibitem[{{Tiede} {et~al.}(2004){Tiede}, {Sarajedini}, \&
  {Barker}}]{2004AJ....128..224T}
{Tiede}, G.~P., {Sarajedini}, A., \& {Barker}, M.~K. 2004, \aj, 128, 224

\bibitem[{{van den Bosch}(2001)}]{2001MNRAS.327.1334V}
{van den Bosch}, F.~C. 2001, \mnras, 327, 1334

\bibitem[{{Westmeier} {et~al.}(2005){Westmeier}, {Braun}, \&
  {Thilker}}]{2005A&A...436..101W}
{Westmeier}, T., {Braun}, R., \& {Thilker}, D. 2005, \aap, 436, 101

\bibitem[{{Westmeier} {et~al.}(2008){Westmeier}, {Br{\"u}ns}, \&
  {Kerp}}]{2008MNRAS.390.1691W}
{Westmeier}, T., {Br{\"u}ns}, C., \& {Kerp}, J. 2008, \mnras, 390, 1691

\bibitem[{{Yagi} {et~al.}(2002){Yagi}, {Kashikawa}, {Sekiguchi}, {Doi},
  {Yasuda}, {Shimasaku}, \& {Okamura}}]{2002AJ....123...66Y}
{Yagi}, M., {Kashikawa}, N., {Sekiguchi}, M., {et~al.} 2002, \aj, 123, 66

\bibitem[{{York} {et~al.}(2000){York}, {Adelman}, {Anderson}, {Anderson},
  {Annis}, {Bahcall}, {Bakken}, {Barkhouser}, {Bastian}, {Berman}, {Boroski},
  {Bracker}, {Briegel}, {Briggs}, {Brinkmann}, {Brunner}, {Burles}, {Carey},
  {Carr}, {Castander}, {Chen}, {Colestock}, {Connolly}, {Crocker}, {Csabai},
  {Czarapata}, {Davis}, {Doi}, {Dombeck}, {Eisenstein}, {Ellman}, {Elms},
  {Evans}, {Fan}, {Federwitz}, {Fiscelli}, {Friedman}, {Frieman}, {Fukugita},
  {Gillespie}, {Gunn}, {Gurbani}, {de Haas}, {Haldeman}, {Harris}, {Hayes},
  {Heckman}, {Hennessy}, {Hindsley}, {Holm}, {Holmgren}, {Huang}, {Hull},
  {Husby}, {Ichikawa}, {Ichikawa}, {Ivezi{\'c}}, {Kent}, {Kim}, {Kinney},
  {Klaene}, {Kleinman}, {Kleinman}, {Knapp}, {Korienek}, {Kron}, {Kunszt},
  {Lamb}, {Lee}, {Leger}, {Limmongkol}, {Lindenmeyer}, {Long}, {Loomis},
  {Loveday}, {Lucinio}, {Lupton}, {MacKinnon}, {Mannery}, {Mantsch}, {Margon},
  {McGehee}, {McKay}, {Meiksin}, {Merelli}, {Monet}, {Munn}, {Narayanan},
  {Nash}, {Neilsen}, {Neswold}, {Newberg}, {Nichol}, {Nicinski}, {Nonino},
  {Okada}, {Okamura}, {Ostriker}, {Owen}, {Pauls}, {Peoples}, {Peterson},
  {Petravick}, {Pier}, {Pope}, {Pordes}, {Prosapio}, {Rechenmacher}, {Quinn},
  {Richards}, {Richmond}, {Rivetta}, {Rockosi}, {Ruthmansdorfer}, {Sandford},
  {Schlegel}, {Schneider}, {Sekiguchi}, {Sergey}, {Shimasaku}, {Siegmund},
  {Smee}, {Smith}, {Snedden}, {Stone}, {Stoughton}, {Strauss}, {Stubbs},
  {SubbaRao}, {Szalay}, {Szapudi}, {Szokoly}, {Thakar}, {Tremonti}, {Tucker},
  {Uomoto}, {Vanden Berk}, {Vogeley}, {Waddell}, {Wang}, {Watanabe},
  {Weinberg}, {Yanny}, \& {Yasuda}}]{2000AJ....120.1579Y}
{York}, D.~G., {Adelman}, J., {Anderson}, Jr., J.~E., {et~al.} 2000, \aj, 120,
  1579

\end{thebibliography}

\end{document}